\newcommand{\typeof}{1}

\documentclass[copyright,creativecommons]{eptcs}
\providecommand{\event}{DICE 2010}
\providecommand{\volume}{??}
\providecommand{\anno}{20??}
\providecommand{\firstpage}{1}
\providecommand{\eid}{??}
\usepackage{breakurl}
\usepackage{ifthen}
\newcommand{\condinc}[2]{\ifthenelse{\equal{\typeof}{0}}{#1}{#2}}
\usepackage{ifpdf}
\ifpdf
	\@error{master}{don's use pdflatek!!!}{}
  \stop
\fi
\usepackage[utf8]{inputenc}	
\usepackage{comment}
\usepackage{amsmath,amsfonts,amssymb}
\usepackage{stmaryrd}
\usepackage{cmll}
\usepackage{latexsym}
\usepackage{mathrsfs}  
\usepackage{proof}
\usepackage{subfigure}  
\usepackage{verbatim}	
\usepackage{alltt}	
\usepackage{float}	
\usepackage[usenames]{color} 
\usepackage[modulo]{lineno}

\usepackage{proofnet}

\usepackage{graphicx}

\newcommand{\subst}[2]{\left[{}^{#1}\!/\!{}_{#2}\right]} 
\newcommand{\substm}[4]{\left[{}^{#1}\!/\!{}_{#2} \cdots {}^{#3}\!/\!{}_{#4}\right]} 

\newcommand{\size}[1]{{|#1|}}

\newcommand{\lin}{\multimap}
\newcommand{\N}{{\mathbb N}}    
\newcommand{\Z}{{\mathbb Z}}    
\newcommand{\B}{{\mathbb B}}    
\newcommand{\NC}{{\mathscr C}}   
\newcommand{\ND}{{\mathscr S}}  
\newcommand{\lang}{\langle}
\newcommand{\rang}{\rangle}

\newcommand{\seq}[3]{\stackrel{\rightarrow}{#1}}  
\newcommand{\numnor}[1]{\mathsf{#1}} 
\newcommand{\numsafe}[1]{\mathsf{#1}} 
\newcommand{\Zero}{\mathtt{0}}               
\newcommand{\Sucz}{\mathtt{s}_0}             
\newcommand{\Suco}{\mathtt{s}_1}             
\newcommand{\Suci}{\mathtt{s}_i}             
\newcommand{\Sucg}[1]{\mathtt{s}_{#1}}             
\newcommand{\Pred}{\mathtt{p}}               
\newcommand{\Proj}[2]{\pi_{#1}^{#2}}         
\newcommand{\Bran}{\mathtt{B}}               
\newcommand{\Rec}[1]{\mathtt{r}[#1]}         
\newcommand{\Comp}[1]{\circ[#1]}             
\newcommand{\degree}[1]{\partial(#1)}        
\newcommand{\strcmplx}{\mathtt{cmplx}}        
\newcommand{\degreeR}[1]{\partial_{\texttt{R}}(#1)}        
\newcommand{\degreeC}[1]{\partial_{\texttt{C}}(#1)}        
\newcommand{\ob}[2]{\textit{tb}_{#1}(#2)}  
\newcommand{\nb}[2]{\textit{nb}_{#1}(#2)}  

\DeclareMathOperator{\pas}{\S}
\DeclareMathOperator{\pa}{\S}
\DeclareMathOperator{\ocs}{!}
\newcommand{\CL}{{\mathcal L}}
\newcommand{\CR}{{\mathcal R}}
\DeclareMathOperator{\cuts}{cuts}

\newcommand{\declareCaption}[2]{\newcommand{#1}{\textnormal{\textbf{#2}}}}

\newcommand{\ALTQ}[2]{%
                      \textnormal{\textbf{$\Lambda_{#1}^{#2}$}}%
                     }
\declareCaption{\MS}{MS}
\declareCaption{\GenDS}{$\mathcal{D}$}
\declareCaption{\PDS}{PDS}
\declareCaption{\LALL}{LALL}
\declareCaption{\FLang}{$\mathcal F$}
\declareCaption{\FLangQ}{${\mathcal F}_{\ND}$}
\declareCaption{\ILAL}{ILAL}
\declareCaption{\LAL}{ILAL}
\declareCaption{\SOLAL}{soLAL}
\declareCaption{\MALL}{MALL}
\declareCaption{\EAL}{IEAL}
\declareCaption{\ELL}{ELL}
\declareCaption{\LLL}{LLL}
\declareCaption{\SLL}{SLL}
\declareCaption{\MLL}{MLL}
\declareCaption{\WALT}{WALT}
\declareCaption{\CS}{CS}
\declareCaption{\IT}{IT}
\declareCaption{\name}{PMS}
\declareCaption{\namebig}{MS}
\declareCaption{\PMS}{PMS}
\declareCaption{\DLAL}{DLAL}
\declareCaption{\LL}{LL}
\declareCaption{\ICC}{ICC}
\declareCaption{\LK}{LK}
\declareCaption{\LJ}{LJ}
\declareCaption{\NK}{NK}
\declareCaption{\NJ}{NJ}
\declareCaption{\PTIME}{PTIME}
\declareCaption{\DTIME}{DTIME}
\declareCaption{\PSPACE}{PSPACE}
\declareCaption{\LOGSPACE}{LOGSPACE}
\declareCaption{\BC}{BC}
\declareCaption{\BCminus}{BC$^-$}
\declareCaption{\BCpm}{BC$^\pm$}
\declareCaption{\SRN}{SRN}
\declareCaption{\MLfour}{ML$^4$}
\declareCaption{\CC}{CC}

\newcommand{\srntolall}[2]{\lceil #1 \rceil^{#2}} 
\newcommand{\srntolallg}{\srntolall{\cdot}{\cdot}}   
\newcommand{\ncof}[1]{\overline{#1}}
\newcommand{\nsof}[1]{[#1]}

\newcommand{\TT}{\texttt{T}}       
\newcommand{\FF}{\texttt{F}}       

\newcommand{\ZW}{\varepsilon\texttt{C}}       
\newcommand{\SW}{\texttt{SuccC}}       
\newcommand{\RevW}{\texttt{RevC}}            

\newcommand{\BoolDiag}{\nabla\texttt{B}}   

\newcommand{\DtoW}{\texttt{StoC}}      
\newcommand{\ZD}{\varepsilon\texttt{S}}       
\newcommand{\SD}{\texttt{SuccS}}       
\newcommand{\PD}{\texttt{PredS}}       
\newcommand{\PPD}{\texttt{PrepS}}       
\newcommand{\BD}{\texttt{CondS}}       
\newcommand{\DD}{\nabla\texttt{S}}       
\newcommand{\DDg}[1]{\DD^{#1}}           
\newcommand{\CoerceD}{\texttt{CoerS}}       
\newcommand{\CD}{\nabla}      

\newcommand{\Lift}[2]{\mathtt{Lift}_{#2}}      

\newcommand{\PTS}{\PTIME-sound}
\newcommand{\pn}{net}
\newcommand{\Pn}{Net}
\newcommand{\formulae}{formul\ae}

\newcommand{\daemon}{d\ae mon}
\newcommand{\Daemon}{D\ae mon}

\DeclareMathOperator{\bbydef}{::=}

\newtheorem{theorem}{Theorem}

\newtheorem{proposition}[theorem]{Proposition}

\newtheorem{definition}[theorem]{Definition}
\newtheorem{fact}[theorem]{Fact}
\newtheorem{remark}[theorem]{Remark}

\newcommand{\titoloblocchi}[1]{#1} 
\newcommand{\capoverso}[1]{\vspace{0.1cm}\par\textbf{#1}}
\makeatletter
\let\adaptedsection\section
\makeatother

\newcommand{\complexfigure}[4][]{
 \ifthenelse{\isempty{#1}}
 {
	\begin{figure}[!tb]
	\begin{center}
	\fbox{
	\begin{minipage}{.9\textwidth}
	#4
	\end{minipage}
	}
	\end{center}
	\caption{#2}
	\label{#3}
	\end{figure}
 }
 {
	\begin{figure}[#1]
	\begin{center}
	\fbox{
	\begin{minipage}{.9\textwidth}
	#4
	\end{minipage}
	}
	\end{center}
	\caption{#2}
	\label{#3}
	\end{figure}
 }
 }

\newcounter{numberone}

\let\oldparagraph\paragraph
\renewcommand{\paragraph}[1]{\oldparagraph{\textnormal{\textbf{#1}}}}

\renewcommand{\vec}[1]{\overrightarrow{#1}}	

\title{Safe Recursion on Notation into a Light Logic by Levels\thanks{Partially supported by MIUR PRIN CONCERTO ---  protocol number 2007BHXCFH}}
\author{Luca Roversi
 \institute{Dipartimento di Informatica -
   Universit\`a di Torino
   }
 \email{roversi@di.unito.it}
 \and
 Luca Vercelli
 \institute{Dipartimento di Matematica -
   Universit\`a di Torino
   }
 \email{luca.vercelli@unito.it}
 }
\def\titlerunning{Safe Recursion on Notation into a Light Logic}
\def\authorrunning{L. Roversi \& L. Vercelli}

\begin{document}
\maketitle

\begin{abstract}
\noindent
We embed Safe Recursion on Notation (\SRN) into Light Affine Logic by Levels (\LALL), derived from the logic \MLfour.
\LALL\ is an intuitionistic deductive system, with a polynomial time cut elimination strategy.
The embedding allows to represent every term $t$ of \SRN\ as a family of \pn s $\lang \srntolall{t}{l}\rang_{l\in\N}$ in \LALL.
Every \pn\ $\srntolall{t}{l}$ in the family simulates $t$ on arguments whose bit length is bounded by the integer $l$.
The embedding is based on two crucial features.
One is the recursive type in \LALL\ that encodes Scott binary numerals, \textit{i.e.} Scott words, as \pn s.
Scott words represent the arguments of $t$ in place of the more standard Church binary numerals.
Also, the embedding exploits the ``fuzzy'' borders of paragraph boxes that \LALL\ inherits from \MLfour\ to ``freely'' duplicate the arguments, especially the safe ones, of $t$.
Finally, the type of $\srntolall{t}{l}$ depends on the number of composition and recursion schemes used to define $t$, namely the structural complexity of $t$. Moreover, the size of $\srntolall{t}{l}$ is a polynomial in $l$, whose degree depends on the structural complexity of $t$.
So, this work makes closer both the predicative recursive theoretic principles \SRN\ relies on, and the proof theoretic one, called \emph{stratification}, at the base of Light Linear Logic.
\end{abstract}

\maketitle
\adaptedsection{Introduction}
\label{section:Introduction}
Slightly rephrasing the \textit{incipit} of \cite{Leivant94ic}, comparing implicit characterizations of computational complexity classes may
provide insights into their nature, while offering concepts and methods for generalizing computational complexity to computing over arbitrary structures and to higher type functions.
Here, we relate two implicit characterizations of polynomial time functions (\PTIME).
One is Safe Recursion on Notation (\SRN) \cite{BellantoniC92}, that we take as representative of the characterizations of \PTIME\ that restrict the primitive recursion. The other one is Light Affine Linear Logic by Levels (\LALL), a proof theoretical system we derive from Light Linear Logic by Levels (\MLfour) \cite{BaillotMazza09} and from Intuitionistic Light Affine Logic (\ILAL) \cite{AspertiR02}. We recall, \MLfour\ and \ILAL\ are two \emph{Light Logics}, \textit{i.e.} restrictions of Linear Logic that characterize some complexity class, in this case \PTIME, under the proofs-as-programs analogy.
These two logics control the complexity of the algorithms they can express by the technical notion \emph{Stratification}, which expresses specific structural restrictions on the derivations of \MLfour\ and \ILAL.
\SRN, of which we recall some more aspects in Section~\ref{section:SRN intermezzo}, provides a \emph{predicative analysis} of primitive recursion. It is the least set that contains
the \emph{zero} $\Zero$ (considered as a 0-ary function),
the \emph{successors} $\Sucz(;x) =  2x, \Suco(;x) = 2x+1$,
the \emph{predecessor} $\Pred(;2x+i) = x$,
the \emph{projection} $\Proj{k}{\numnor n;\numnor s}(\seq{x}{1}{\numnor{n}};\seq{y}{1}{\numsafe{s}}) = x_k$ if $1\leq k \leq\numnor n$, and $y_{k}$ if $1 \leq  k \leq \numsafe{s}$,
the \emph{conditional} $\Bran(;y,y_1,y_2)= y_1$ if $y$ is odd, and $y_2$ otherwise,
and which is closed under \emph{safe composition}  and \emph{predicative recursion on notation} (\eqref{some-SRN:composition} and \eqref{some-SRN:recursion} in Figure~\ref{fig:some-SRN}).
The work \cite{MurawskiOng00} is the first one relating the two different traditions:
it defines a map from terms of a strict fragment \BCminus\ of \SRN\ into \pn s of \ILAL.
The main obstacle to a full representation of \SRN\ into \LAL\ is that the duplication of \pn s in \ILAL, hence of the  safe arguments, is far from being free, as required instead by \eqref{some-SRN:recursion}.
In fact, \cite{MurawskiOng00} also shows that an extension \BCpm, polynomial time complete, can be represented inside \ILAL.
However, since the primitives added to \BCpm\ are not in \SRN, we cannot see \BCpm\ as relevant to the goal of understanding the possible relation between full \SRN\ and the above stratification principle, basic to \MLfour\ and \ILAL.
Since \cite{MurawskiOng00}, no extension of the relation between \SRN\ and \LAL\ has been produced, to our knowledge.
Here, we show to which extent we can avoid that obstacle inside \LALL.
\LALL, that will be formally defined in Section~\ref{section:Definitions and Basic Properties tempname},
is an intuitionistic system of \pn s endowed with:
(i) edges labelled by indices, or levels,
(ii) unconstrained weakening, to make programming with its \pn s somewhat more comfortable,
(iii) a language of \formulae\ quotiented by the recursive equivalence
$\ND = \forall\alpha.(\alpha\multimap(\B\lin \ND \multimap\alpha)\multimap\alpha)$, where $\B$ is the type of booleans,
and $\ND$ the data type of \emph{Scott words} \cite{AbadiCardelliPlotkin}, and
(iv) a polynomial time sound cut elimination procedure (Section~\ref{section:Proof of Polynomial Soundness for tempname}) which does not depend on the types that label the edges of a given \pn.
\par
\SRN\ embeds into \LALL\ by means of the map $\srntolallg$ (Sections~\ref{section:Programming inside tempname} and \ref{section:Defining the embedding srntolall from SRN to LALL}.)
The map $\srntolallg$ has the same \emph{natural} and \emph{inductive} structure as the one of the map in \cite{MurawskiOng00} from \SRN\ to \LAL.
However, $\srntolallg$ takes two arguments:
(i) any term $t$ of \SRN${}^{\numnor{n};\numsafe{s}}$, with normal arity $\numnor{n}$, the number of arguments to the left of the semicolon, and safe arity $\numsafe{s}$, the number of those ones to the right, and
(ii) an integer $l\geq 0$ that bounds the size of every argument of $t$.
Then, $\srntolall{t}{l}$ yields a \pn\ that simulates $t(\vec x;\vec y)$ whenever every element in $\vec x,\vec y$ \emph{is at most as long as} $l$ (Proposition~\ref{proposition:tempname simulates SRN}).
This suggests to summarize the situation we move in by:
{
\begin{align}
\label{align:analogy}
\frac{\textit{\LALL}}{\SRN} =
\frac{\textrm{\PTIME-uniform Boolean Circuits}}{\textrm{\PTIME\ Problems}}
\end{align}
}
\noindent
We remark, however, that such an analogy should be read as such, and not as a formal correspondence.
\textit{I.e.}, we are not at all assuming any classical complexity theoretic perspective like the one in \cite{DBLP:conf/lics/Terui04}, which shows a proofs-as-programs correspondence between Boolean Circuits and \pn s of Multiplicative Linear Logic.
\par
Instead, what we do reads as follows.
\par
Let $t$ be a term of \SRN${}^{\numnor{n};\numsafe{s}}$.
We write $\degreeC{t}$ and $\degreeR{t}$ for the number of composition and recursion schemes, respectively, that are used to build $t$.
That way, $\strcmplx(t)=\degreeC{t}+\degreeR{t}$ is a na\"ive measure of the \emph{static} complexity of $t$.
Also, let $p_t$ be the characteristic polynomial of $t$, whose values bound the length of the output of $t$.
Let $\degree{p_t}$ be its degree. Then, $t$ is represented in \LALL\ by a family $\lang \srntolall{t}{l} \rang_{l\in\N}$ of \pn s such that:
\begin{enumerate}
\item
The size of every \pn\ $\srntolall{t}{l}$ is $O(l^{\degree{p_t}^{\strcmplx(t)}})$, namely polynomial in $l$;

\item
If $l$ is at least as great as every bit length
$\size{x_1},\ldots,\size{x_\numnor{n}},\size{y_1},\ldots,\size{y_\numsafe{s}}$ of the arguments, then
the application of $\srntolall{t}{l}$ to
$\srntolall{x_1}{l},\ldots,\srntolall{x_\numnor{n}}{l},
 \srntolall{y_1}{l},\ldots,\srntolall{y_\numsafe{s}}{l}$
equals
$\srntolall{t(x_1,\ldots,x_{\numnor{n}};y_1,\ldots,y_{\numsafe{s}})}{l}$
;

\item
Every $\srntolall{t}{l}$ is a map from
$(\bigotimes^{\numnor{n}}\ND)\otimes (\bigotimes^{\numsafe{s}}(\pas^k\ND))$ to $\pas^k\ND$,
where $k$ depends on $\strcmplx(t)$.
\end{enumerate}
\noindent
The first two points suggest the analogy \eqref{align:analogy}. Specifically, the first point expresses a uniformity condition on the \pn s in the family, since it states that their dimension are bounded only by the length of the inputs. The second point says that $\srntolall{t}{l}$ soundly simulates $t$ on \emph{every} input of length \emph{smaller or equal} to $l$.
Finally, the third point is a natural property we can expect as soon as we try to compositionally and naturally represent first order algebraic terms, that operate on a given domain, into a higher order language. It is a static description of the behavior of $t$ in terms of types of \LALL, a kind of information we cannot have by, for example, representing \SRN\ as circuit families.
\par
We see the use of $\ND$ as a first fundamental choice to write $\srntolallg$.
The reason is twofold.
One reason is a kind of obvious, since $\ND$ supports the representation of successors, predecessor, projection and conditional as constant time operations, unlike
$\NC=\forall\alpha.\oc(\alpha\multimap\alpha)\multimap\oc(\alpha\multimap\alpha)\multimap\pa(\alpha\multimap\alpha)$, which is generally used to represent \emph{Church words} in Light Logics.
The other reason, instead, brings a certain degree of novelty with it because we exploit a crucial property of \LALL, and of Light Logics, \emph{which had hardly been used so far}.
The crucial property is that the polynomial time cut elimination of \LALL\ depends only on the structure of any given \pn\ $\Pi$, while the logical complexity of the \formulae\ in $\Pi$ does not affect it.
So, we are free to add fixpoints \formulae, like $\ND$ is, which adds a huge expressivity to the logic.
\par
A second step to get $\srntolallg$, for every $l$, we exploit what we like to call the \emph{fuzzy} borders of paragraph boxes of \LALL\ to write the \pn\ $\DDg{k}_l$. The \pn\ $\DDg{k}_l$ duplicates a Scott word at most as long as $l$, starting from a premise of type $\S^k\ND$ and concluding with the type  $\S^k\ND\otimes\S^k\ND$, for any $k$.
We remark that in \LAL, where the border of paragraph boxes is ``rigid'', we could only write a \pn, analogous to $\DDg{k}_l$, concluding with type $\S^k(\ND\otimes\ND)$ which would generally impede to get the right type for $\srntolall{t}{l}$.
By the way, this is why \cite{MurawskiOng00} shows how to embed \BCminus\ but not \SRN\ into \LAL.
Indeed, in \BCminus, composition and safe recursion schemes allow \emph{linear} safe arguments only, i.e. the safe arguments are never duplicated.
\par
To conclude, we recall what \emph{stratification} means. It is a structural property underpinning the \PTS\ cut elimination of Girard's Light Linear Logic (\LLL) \cite{Girard98}, and its variants \LAL, \MLfour, and \LALL.
A \pn\ $\Pi$ is \textit{stratified} if the number of boxes around every node keeps being constant in every \pn\ we reach from $\Pi$ by cut elimination.
This work should be a step further towards studying how the \emph{stratification} is compatible with the predicative analysis of \PTS\ computations that \SRN\ embodies.
\complexfigure[t]
{\SRN: predicative recursion on notation and safe composition.}
{fig:some-SRN}
{
\small%
 \begin{align}%
\label{some-SRN:composition}
\Comp{t'\,u_1\ldots u_{\numnor{n}'}\,
          v_1\ldots v_{\numsafe{s}'}
      }(\seq{x}{1}{\numnor{n}};\seq{y}{1}{\numsafe{s}})
&=
t'(u_1            (\seq{x}{1}{\numnor{n}};)
  ,\ldots,
  u_{\numnor{n}'}(\seq{x}{1}{\numnor{n}};)
  ;
  v_1(\seq{x}{1}{\numnor{n}};\seq{y}{1}{\numsafe{s}})
  ,\ldots,
  v_{\numsafe{s}'}(\seq{x}{1}{\numnor{n}};\seq{y}{1}{\numsafe{s}}))
\\
\label{some-SRN:recursion}
 \nVoid{anchor1}
 \Rec{u_\varepsilon,u_0,u_1}
 (\Zero,\seq{x}{1}{\numnor{n}};\seq{y}{1}{\numsafe{s}})
 &= u_\varepsilon(\seq{x}{1}{\numnor{n}};\seq{y}{1}{\numsafe{s}})
 \\
\nonumber
 \Rec{u_\varepsilon,u_0,u_1}
 (2z+i,\seq{x}{1}{\numnor{n}};\seq{y}{1}{\numsafe{s}})
 &=u_i(z,\seq{x}{1}{\numnor{n}};\seq{y}{1}{\numsafe{s}}
      , \Rec{u_\varepsilon,u_0,u_1}
        (z,\seq{x}{1}{\numnor{n}};\seq{y}{1}{\numsafe{s}}))
      &i\in\{0,1\}
\end{align}
 \uput{0.8cm}[195](anchor1){\scalebox{2}[4]{\{}}
}
\paragraph{Acknowledgments.} We want to thank the anonymous referees whose questions helped us to better address the points subject of this work.
\complexfigure
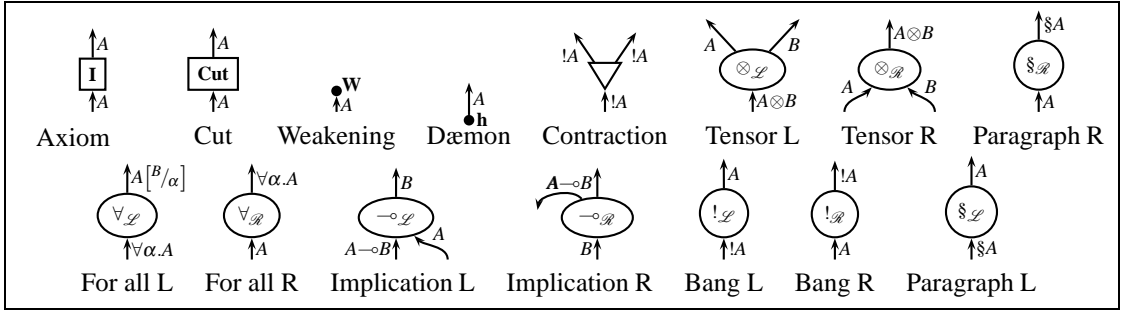
{The nodes in the proof nets of \LALL.}
{fig:ilal}
{
\psset{colsep=5pt} 
\small
\centering
\begin{tabular}{cccccccccc}
  \proofnet{
   \nVoid{out}\\
   \nId{id}\\
   \nVoid{in}
   \ncline{in}{id} \nbput A
   \ncline{id}{out} \nbput A
  }
&
  \proofnet{
   \nVoid{out}\\
   \nCut{ct}\\
   \nVoid{in}
   \ncline{in}{ct} \nbput A
   \ncline{ct}{out} \nbput A
  }
&
  \proofnet{
   \nWea{w}\\
   \nVoid{in}
   \ncline{in}{w} \nbput A
  }
&
  \proofnet{
   \nVoid{out}\\
   \nDaem{h}
   \ncline{h}{out} \nbput A
  }
&
  \proofnet{
   \nVoid{o1}&&  \nVoid{o2}\\
   &\nCon{y}\\
   &\nVoid{in}
   \ncline[offsetA=1pt]{y}{o1} \naput{\oc A}
   \ncline[offsetA=-1pt]{y}{o2} \nbput{\oc A}
   \ncline{in}{y} \nbput{\oc A}
  }
&
  \proofnet{
   \nVoid{o1}&&\nVoid{o2}\\
   &\nLtens{t}\\
   &\nVoid{in}
   \ncline{in}{t} \nbput{A\otimes B}
   \ncline[angleA=135]{t}{o1} \naput A
   \ncline[angleA=45]{t}{o2} \nbput B
  }
&
  \proofnet{
   &\nVoid{o}\\
   &\nRtens{t}\\
   \nVoid{i1}&&\nVoid{i2}
   \nccurve[angleA=90,angleB=-135]{i1}{t} \naput A
   \nccurve[angleA=90,angleB=-45]{i2}{t} \nbput B
   \ncline{t}{o} \nbput{A\otimes B}
  }
&
  \proofnet{
   \nVoid{out}\\
   \nRPA{pax}\\
   \nVoid{in}
   \ncline{in}{pax} \nbput{A}
   \ncline{pax}{out} \nbput{\S A}
  }
\\
\begin{minipage}{1.5cm}Axiom\end{minipage}&
Cut&
Weakening&
\Daemon&
Contraction&
Tensor L&
Tensor R&
Paragraph R
\end{tabular}

\begin{tabular}{ccccccccc}
  \proofnet{
   \nVoid{c}\\
   \nLfor{b}\\
   \nVoid{a}
   \ncline{a}{b} \nbput{\forall\alpha.A}
   \ncline{b}{c} \nbput{A\subst B\alpha}
  }
&
  \proofnet{
   \nVoid{c}\\
   \nRfor{b}\\
   \nVoid{a}
   \ncline{a}{b} \nbput{A}
   \ncline{b}{c} \nbput{\forall\alpha.A}
  }
&
  \proofnet{
   \nVoid{o}\\
   \nLlin{imp}\\
   \nVoid{i1}&\nVoid{i2}
   \ncline{imp}{o} \nbput B
   \ncline{i1}{imp} \naput{A\multimap B}
   \nccurve[angleA=90,angleB=-45]{i2}{imp} \nbput{A}
  }
&
  \proofnet{
   &\nVoid{out}\\
   \rnode{h}{\phantom\multimap}&\nRlin{imp}\\
   &\nVoid{in}
   \ncline{in}{imp} \naput B
   \ncline{imp}{out} \naput{A\multimap B}
   \nccurve[angleA=130,angleB=80]{imp}{h} \nbput{A}
  }
&
  \proofnet{
   \nVoid{out}\\
   \nLOC{p}\\
   \nVoid{in}
   \ncline{in}{p} \nbput{\oc A}
   \ncline{p}{out}  \nbput{A}
  }
&
  \proofnet{
   \nVoid{out}\\
   \nROC{pax}\\
   \nVoid{in}
   \ncline{in}{pax} \nbput{A}
   \ncline{pax}{out} \nbput{\oc A}
  }
&
  \proofnet{
   \nVoid{out}\\
   \nLPA{p}\\
   \nVoid{in}
   \ncline{in}{p} \nbput{\S A}
   \ncline{p}{out}  \nbput{A}
  }
\\
For all L&
For all R&
Implication L&
Implication R&
Bang L&
Bang R &
Paragraph L\end{tabular}
} 

\adaptedsection{Light Affine Logic by Levels (\LALL)}
\label{section:Definitions and Basic Properties tempname}
\capoverso{The language of \formulae.}
First, for any fixed countable set $\mathcal V$ of propositional variables, the set \FLang\ of \formulae\ is generated by the following grammar:
$$F\bbydef \ND \mid \alpha\mid F\otimes F\mid F\multimap F\mid \forall\alpha.F
        \mid\ocs F\mid\pas F\qquad\alpha\in\mathcal V$$
where $\ND$ is a propositional constant.
Second, we define the quotient \FLangQ\ of \FLang\ by assuming:
{
\small
\begin{align}
\label{eqref:equivalence-on-ND}
\ND & = \forall\alpha.(\alpha\multimap(\B\lin \ND \multimap\alpha)\multimap\alpha)
\end{align}
}
among the elements of \FLang. Namely, \eqref{eqref:equivalence-on-ND} says that $\ND$ represents Scott words \cite{AbadiCardelliPlotkin}.
The \formulae\ we shall effectively use are the equivalence classes in \FLangQ. Every time we label an edge of a \pn\ of \LALL\ by $\ND$, we can also label that edge by any ``unfolding'' of $\ND$ that obeys \eqref{eqref:equivalence-on-ND}.
$A\subst{B}{\alpha}$ is the substitution of every free occurrence of $\alpha$ in $A$ with $B$.
\capoverso{Proof structures and \pn s.}
\LALL\ is a language of \pn s. \Pn s will be defined as particular proof structures.
Given the nodes in Figure~\ref{fig:ilal}, we say that \emph{an Axiom node and a \Daemon\ nodes are proof structures}.
Moreover, given two proof structures $\Pi$ and $\Sigma$:
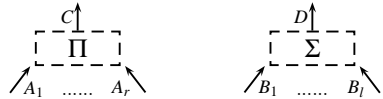
\begin{center}
\begin{tabular}{ccc}
\inlineproofnet[rowsep=.3,colsep=.2]{0.9}{
\nVoid{out} \\
\simplenet{a}{\Pi} \\
\rnode{b}{\ldots\ldots}
\ncline[offsetA=1,offsetB=.7]{b}{a}
  \nbput{A_1}
\ncline[offsetA=-1,offsetB=-.7]{b}{a}
  \naput{A_ r}
\ncline{a}{out} \naput{C}
}
&\qquad\qquad&
\inlineproofnet[rowsep=.3,colsep=.2]{0.9}{
\nVoid{out} \\
\simplenet{a}{\Sigma} \\
\rnode{b}{\ldots\ldots}
\ncline[offsetA=1,offsetB=.7]{b}{a}
  \nbput{B_1}
\ncline[offsetA=-1,offsetB=-.7]{b}{a}
  \naput{B_l}
\ncline{a}{out} \naput{D}
}
\end{tabular}
\end{center}
denoted as ${\Pi}\triangleright{A_1,\ldots,A_r}\vdash C$ and
${\Sigma}\triangleright{B_1,\ldots,B_l}\vdash D$, respectively,
with $r,l \geq 0$, then all the graphs inductively built from
$\Pi$ and $\Sigma$ by the rule schemes in Figure~\ref{fig:proof-nets-ILAL} are proof structures.
\par
If $\Pi\triangleright\Gamma\vdash A$, we say that $\Pi$ \emph{proves}
the sequent $ \Gamma\vdash A$. The \emph{inputs} (resp. \emph{outputs}) of $\Pi$ are the edges labelled
$\Gamma$ (resp. $A$).
The set of the nodes of $\Pi$ is $V_\Pi$, and $E_\Pi$ is the set of edges.
The \emph{size} $|\Pi|$ of $\Pi$ is the cardinality of $V_\Pi$.
The \emph{depth} $\partial(x)$ of a node or edge $x\in V_\Pi\cup E_\Pi$ is the number of nested $\ocs$-boxes containing $x$.
The \emph{depth} $\partial(\Pi)$ of $\Pi$ is the greatest depth among the nodes of $\Pi$.
\par
Every $!$-box simultaneously introduces one Bang R node and at most one Bang L node, recording this by the box border as in Figure~\ref{fig:proof-nets-ILAL}.
\complexfigure
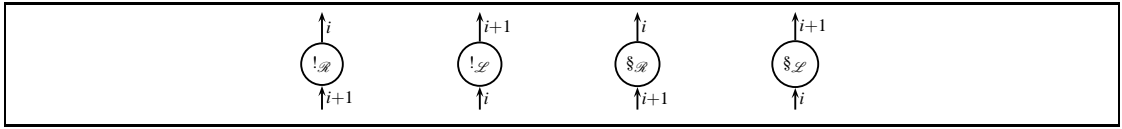
{Constraints on the indexing.
The nodes we omit have the same index on all of their incident edges.}
{fig:constraints-indexing-mell}
{
\small
\centering
\begin{tabular}{cccccccc}
\proofnetpagescaled[rowsep=.3,colsep=.2]{0.9}{1.5cm}{
\nVoid{a} \\
\nROC{b} \\
\nVoid{c}
\ncline{b}{a}
  \nbput{i}
\ncline{c}{b}
  \nbput{i+1}
}
&
\proofnetpagescaled[rowsep=.3,colsep=.2]{0.9}{1.5cm}{
\nVoid{a} \\
\nLOC{b} \\
\nVoid{c}
\ncline{b}{a}
  \nbput{i+1}
\ncline{c}{b}
  \nbput{i}
}
&
\proofnetpagescaled[rowsep=.3,colsep=.2]{0.9}{1.5cm}{
\nVoid{a} \\
\nRPA{b} \\
\nVoid{c}
\ncline{b}{a}
  \nbput{i}
\ncline{c}{b}
  \nbput{i+1}
}
&
\proofnetpagescaled[rowsep=.3,colsep=.2]{0.9}{1.5cm}{
\nVoid{a} \\
\nLPA{b} \\
\nVoid{c}
\ncline{b}{a}
  \nbput{i+1}
\ncline{c}{b}
  \nbput{i}
}
\end{tabular}
}
\begin{definition}[\titoloblocchi{Indexing and \Pn s, adapted from \cite{BaillotMazza09}}] \label{def:index-ml3}
 Let $\Pi$ be a proof structure.
\begin{enumerate}
 \item
  An \emph{indexing} for $\Pi$ is a function $I$ from the edges of $\Pi$ to $\Z$ that
  satisfies the constraints in Figure~\ref{fig:constraints-indexing-mell}
  and such that $I(e)=I(e')$, for every pair $e,e'$ of inputs and output of $\Pi$.
 \item
  A \emph{\pn} is a proof structure that admits an indexing.
 \item
  An indexing $I$ of $\Pi$ is \emph{canonical} if $\Pi$
  has an edge $e$ such that $I(e) = 0$, and $I (e') \geq 0$ for all edges $e'$ of $\Pi$.
\end{enumerate}
\end{definition}
As in \cite{BaillotMazza09}, we can state that every \pn\ of \LALL\ admits a unique canonical indexing.
\par
The indexing tells that the nodes $!_{\mathcal L}$ and $\S_{\mathcal L}$ are not \emph{dereliction} nodes.
Remember that the dereliction rule of Linear Logic is inherently not stratified, because the cut-elimination is presence of a dereliction node may also ``open'' boxes.
Instead, these nodes can be considered as auxiliary ports of $\S$-boxes whose border is somewhat \emph{fuzzy}.
We mean that a $\S$-box need not be contained in or disjoint from another box.
Instead, it can ``overlap'' a $!$-box, and it can have more than one conclusion $\pas_\CR$.
To distinguish $\S$-boxes from the $!$ ones we adopt a dotted border.
\par
Let $I_0$ be the canonical indexing of $\Pi$ and $e\in E_\Pi$.
The \emph{level of $e$} is $l(e)$. It is defined as $I_0(e)$.
The \emph{level of $\Pi$} is $l(\Pi)$. It is defined as the greatest value assumed by $I_0$ on the edges of $\Pi$.
We denote as $B_\Pi$ the set of the $\ocs$-boxes in $\Pi$, and it is naturally in bijection with the set of the $\ocs_\CR$ nodes in $\Pi$.
Finally, for every \pn\ $\Pi$, and for $\dagger\in\{\ocs,\pas\}$, $\dagger^n\Pi$ denotes $n$ nested $\dagger$-boxes around $\Pi$.
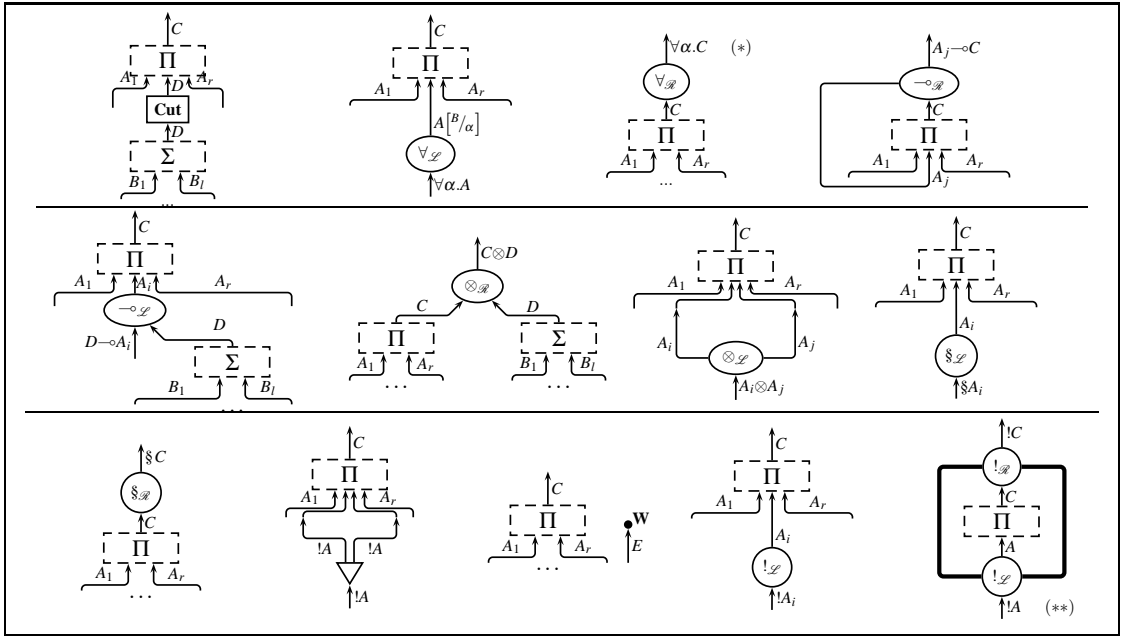
\complexfigure{Inductive rule schemes to build proof structures of \LALL.
(*) $\alpha$ does not occur free in $A_1,\ldots,A_r$.
(**) A $!$-box, which has \emph{at most} a single assumption.
}
{fig:proof-nets-ILAL}
{
        \centering
        \begin{tabular}{cccc}
        \psset{labelsep=1pt}
        \proofnetpagescaled[colsep=4pt]{0.8}{0.2\textwidth}{   
                                &&\nVoid{out}\\
                                &&\simplenet{p}{\Pi}\\
                                \nVoid{p1} &&\nCut{cut} &&\nVoid{p2}\\
                                &&\simplenet{s}{\Sigma}\\
                                &\nVoid{s1} &\rnode{s2}{\ldots} &\nVoid{s3}
                                \ncline{p}{out} \nbput C
                                \ncup[offsetB=10pt,armB=2mm]{p1}{p} \naput{A_1}
                                \ncup[offsetB=-10pt,armB=2mm]{p2}{p} \nbput{A_r}
                                \ncline{cut}{p} \nbput{D}
                                \ncline{s}{cut} \nbput{D}
                                \ncup[offsetB=6pt]{s1}{s} \naput{B_1}
                                \ncup[offsetB=-6pt]{s3}{s} \nbput{B_l}
                }
        &
        \proofnetpagescaled{0.8}{0.2\textwidth}{  
                                &&\nVoid{out}\\
                                &&\simplenet{p}{\Pi}\\
                                \nVoid{i1} && && \nVoid{i3}\\
                                &&\nLfor{w}\\
                                &&\nVoid{i4}
                                \ncline{p}{out} \nbput C
                                \ncup[offsetB=6pt]{i1}{p} \naput{A_1}
                                \ncline{w}{p} \nbput[npos=0.2]{A\subst B \alpha{}}
                                \ncup[offsetB=-6pt]{i3}{p} \nbput{A_r}
                                \ncline{i4}{w} \nbput[labelsep=1pt]{\forall\alpha.A}
                        }
        &
        \proofnetpagescaled{0.8}{0.15\textwidth}{   
                                &\nVoid{out}\\
                                &\nRfor{v}\\
                                &\simplenet{p}{\Pi}\\
                                \nVoid{i1} &\rnode{i2}{\ldots} &\nVoid{i3}
                                \ncline{v}{out} \nbput{\forall\alpha.C\quad(*)}
                                \ncline{p}{v} \nbput C
                                \ncup[offsetB=6pt]{i1}{p} \naput{A_1}
                                \ncup[offsetB=-6pt]{i3}{p} \nbput{A_r}
                        }
        &
        \proofnetpagescaled{0.8}{0.25\textwidth}{  
                                &&\nVoid{out}\\
                                &&\nRlin{v}\\
                                &&\simplenet{p}{\Pi}\\
                                \nVoid{i1} & &&&\nVoid{i2}
                                \ncline{v}{out} \nbput{A_j\multimap C}
                                \ncline{p}{v} \nbput C
                                \ncup[offsetB=6pt]{i1}{p} \naput{A_1}
                                \ncline{kk}{p}
                                \ncup[offsetB=-6pt]{i2}{p} \nbput{A_r}
                                \ncldownup[armA=13mm,armB=6mm]{v}{p} \nbput[npos=3.2]{A_j}
                        }
        \\ \hline
        \end{tabular}\\
        \begin{tabular}{cccc}
        \proofnetpagescaled{0.8}{0.23\textwidth}{  
                                &&\nVoid{out}\\
                                &&\simplenet{p}{\Pi}\\
                                \nVoid{i1} &&\nLlin{v} &&\nVoid{i2}\\
                                &&\nVoid{i3}&\simplenet{s}{\Sigma}\\
                                &&\nVoid{i4} &\ldots &\nVoid{i6}
                                \ncline{p}{out} \nbput C
                                \ncup[offsetB=10pt,armB=3mm]{i1}{p} \naput{A_1}
                                \ncup[offsetB=-10pt,armB=3mm]{i2}{p} \nbput{A_r}
                                \ncline{v}{p} \nbput[labelsep=0.1pt]{A_i}
                                \ncline{i3}{v} \naput{D\multimap A_i}
                                \ncup[angleB=-45]{s}{v} \uput{.5cm}[110](s){D}
                                \ncup[offsetB=6pt]{i4}{s} \naput{B_1}
                                \ncup[offsetB=-6pt]{i6}{s} \nbput{B_l}
                }
        &
        \proofnetpagescaled[colsep=4pt]{0.8}{0.22\textwidth}{   
                &&&\nVoid{out}\\
                                &&&\nRtens{tr}\\
                                &\simplenet{p}{\Pi}&&&&\simplenet{s}{\Sigma}\\
                                \nVoid{p1} &\ldots &\nVoid{p3} &&\nVoid{s1} &\ldots &\nVoid{s3}
                                \ncline{tr}{out} \nbput{C\otimes D}
                                \ncup[angleB=-135]{p}{tr} \naput C
                                \ncup[angleB=-45]{s}{tr} \nbput D
                                \ncup[offsetB=6pt]{p1}{p} \naput{A_1}
                                \ncup[offsetB=-6pt]{p3}{p} \nbput{A_r}
                                \ncup[offsetB=6pt]{s1}{s} \naput{B_1}
                                \ncup[offsetB=-6pt]{s3}{s} \nbput{B_l}
                        }
        &
        \proofnetpagescaled{0.8}{0.15\textwidth}{   
                                &&&\nVoid{out}\\
                                &&&\simplenet{p}{\Pi}\\
                                \nVoid{i1} &&\nVoid{p1} &&\nVoid{p2} &&\nVoid{i2}\\
                                &&&\nLtens{v}\\
                                &&&\nVoid{i3}
                                \ncline{p}{out} \nbput C
                                \ncup[offsetB=7pt,armB=2mm]{i1}{p} \naput{A_1}
                                \ncup[armB=3mm,offsetB=2pt]{p1}{p}
                                \ncup[armB=3mm,offsetB=-2pt]{p2}{p}
                                \ncup[offsetB=-7pt,armB=2mm]{i2}{p} \nbput{A_r}
                                \ncangle[angleA=180,angleB=-90]{v}{p1} \naput{A_i}
                                \ncangle[angleA=0,angleB=-90]{v}{p2} \nbput{A_j}
                                \ncline{i3}{v} \nbput{A_i\otimes A_j}
                                }
        &
        \proofnetpagescaled{0.8}{0.22\textwidth}{  
                                &&\nVoid{out}\\
                                &&\simplenet{p}{\Pi}\\
                                \nVoid{i1} && && \nVoid{i3}\\
                                &&\nLPA{w}\\
                                &&\nVoid{i4}
                                \ncline{p}{out} \nbput C
                                \ncup[offsetB=6pt]{i1}{p} \naput{A_1}
                                \ncline{w}{p} \nbput[npos=0.2]{A_i}
                                \ncup[offsetB=-6pt]{i3}{p} \nbput{A_r}
                                \ncline{i4}{w} \nbput[labelsep=1pt]{\S A_i}
                        }
        \\ \hline
        \end{tabular}\\
        \begin{tabular}{ccccc}
        \proofnetpagescaled{0.8}{0.15\textwidth}{ 
                                &\nVoid{out}\\
                                &\nRPA{v}\\
                                &\simplenet{p}{\Pi}\\
                                \nVoid{i1} &\ldots &\nVoid{i3}
                                \ncline{v}{out} \nbput{\pas C}
                                \ncline{p}{v} \nbput{C}
                                \ncup[offsetB=6pt]{i1}{p} \naput{A_1}
                                \ncup[offsetB=-6pt]{i3}{p} \nbput{A_r}
                        }
        &
        \proofnetpagescaled[colsep=4pt]{0.8}{0.15\textwidth}{  
                                &&&\nVoid{out}\\
                                &&&\simplenet{p}{\Pi}\\
                                \nVoid{p1} &&\nVoid{X} &&\nVoid{Y} &&\nVoid{p3}\\
                                &&&\nCon{yy}\\
                                &&&\nVoid{in}
                                \ncline{p}{out} \nbput C
                                \ncup[offsetB=7pt,armB=3mm]{p1}{p} \naput{A_1}
                                \ncup[offsetB=-7pt,armB=3mm]{p3}{p} \nbput{A_r}
                                \ncup[armB=3.5mm,offsetB=2pt]{X}{p}
                                \ncup[armB=3.5mm,offsetB=-2pt]{Y}{p}
                                \ncup[offsetA=2pt]{yy}{X} \naput{\oc A}
                                \ncup[offsetA=-2pt]{yy}{Y} \nbput{\oc A}
                                \ncline{in}{yy} \nbput{\oc A}
                                }
        &
        \proofnetpagescaled{0.8}{0.15\textwidth}{  
                                &\nVoid{out}\\
                                &\simplenet{p}{\Pi}&&\nWea{w}\\
                                \nVoid{p1} &\ldots &\nVoid{p3} &\nVoid{p4}
                                \ncline{p}{out} \nbput C
                                \ncup[offsetB=6pt]{p1}{p} \naput{A_1}
                                \ncup[offsetB=-6pt]{p3}{p} \nbput{A_r}
                                \ncline{p4}{w} \nbput E
        }
        &
        \proofnetpagescaled{0.8}{0.15\textwidth}{  
                                &&\nVoid{out}\\
                                &&\simplenet{p}{\Pi}\\
                                \nVoid{i1} && && \nVoid{i3}\\
                                &&\nLOC{w}\\
                                &&\nVoid{i4}
                                \ncline{p}{out} \nbput C
                                \ncup[offsetB=6pt]{i1}{p} \naput{A_1}
                                \ncline{w}{p} \nbput[npos=0.2]{A_i}
                                \ncup[offsetB=-6pt]{i3}{p} \nbput{A_r}
                                \ncline{i4}{w} \nbput[labelsep=1pt]{!A_i}
                        }
        &
        \proofnetpagescaled{0.8}{0.15\textwidth}{  
                &                   &&\nVoid{out}\\
                &                   &&\nROC{pout}\\
                &       &&\simplenet{p}{\Pi} \\
                &&&\nLOC{pdue} &\\
                &&\nVoid{i1}  &\nVoid{i2}&\nVoid{i3}
        \ncline{pout}{out} \nbput{\oc C}
        \ncline{p}{pout} \nbput C
        \ncline{pdue}{p} \nbput{A}
        \ncline{i2}{pdue}  \nbput{\oc A\quad(**)}
        \abox[arm=7mm]{1}{pout}{pdue}
        }
        \end{tabular}
}

\capoverso{Cut elimination.}
We just recall its steps, which are standard.
The \emph{linear} cut elimination steps annihilate in the natural way a pair of linear nodes
(Identity/Cut, $\lin_\CL/\lin_\CR$, $\otimes_\CL/\otimes_\CR$, $\pas_\CL/\pas_\CR$, $\forall_\CL/\forall_\CR$).
The \emph{exponential} cut elimination steps are of two kinds: $\ocs_\CL/\ocs_\CR$ is reduced merging the two involved boxes
which can be $\ocs$-boxes as well as $\pas$-boxes with fuzzy borders.
Instead, \emph{contraction}/$\ocs_\CR$ duplicates the whole $\ocs$-box cut with the contraction, as in \ILAL.
The \emph{garbage collection} cut elimination steps involve the weakening or the \daemon\ nodes, cut with any other node.
It is always possible to reduce such a cut with the help of some more weakening and \daemon\ nodes,
as done in \ILAL\ \cite{AspertiR02}.
The set of cut nodes of $\Pi$ is $\cuts(\Pi)$.
%
%
\begin{proposition}[Cut-elimination]
 Every \LALL\ \pn\ reduces to a cut-free \pn.
\end{proposition}

A direct proof would be very long; anyway, such a proof directly follows from the proof of the namesake propositions in \ILAL\ and \MLfour. Please notice that the presence of fixpoints (i.e. the recursive type $\ND$) does not affect the proof in any way, because the cut-elimination independently by the \formulae\ labelling the edges of a \pn. This is not true in full Linear Logic.
\adaptedsection{Polynomial time soundness of \LALL}
\label{section:Proof of Polynomial Soundness for tempname}
We adapt \cite{BaillotMazza09} to prove the cut elimination \PTS ness in presence of unconstrained Weakening, which we introduce for easy of programming since it is handy to erase \pn s structure.
Let us fix a proof net $\Pi$ to reduce. We define an ordering over $\cuts(\Pi)$ that determines which cuts to reduce first.\par

A graph theoretic path in any proof net $\Pi$ is \emph{exponential} if it contains a, possibly empty, sequence of consecutive contractions and stops at a $\ocs_\CL$ node.\par

Let $B,C\in B_\Pi$. Let $B\prec_1^L C$ if the roots of $B$ and $C$ lie at the same level, and
the root of $B$ is in cut with an exponential path that enters an auxiliary port of $C$.
$\preceq^L$ is the reflexive and transitive closure of $\prec_1^L$.
One can show that $\preceq^L$ is a partial order, \emph{upward arborescent}:
for every $C$ there is at most one $B$ such that $B\prec_1^L C$.\par

Let $c,c'\in\cuts(\Pi)$.
We write $c\leq c'$ iff one of the following conditions holds.
(i) $c'$ is connected to a weakening or a \daemon, and $c$ is not.
(ii) The condition (i) is false but $l(c)<l(c')$ holds.
(iii) The conditions (i) and (ii) are false, so $l(c)=l(c')$. In this case, $c\leq c'$ iff:
  (a) either $c'$ is connected to a contraction, and $c$ is not, or
  (b) $c, c'$ are connected to a contraction on one side, to the boxes $B, B'$, respectively, on the other, and $B\preceq^L B'$.

\begin{definition}[\titoloblocchi{Canonical normalization}]
\label{definition:Canonical normalization}
A sequence of normalization steps that starts from a given proof net $\Pi$ is \emph{canonical} whenever smaller cuts relatively to $\leq$ are eliminated before higher ones.
\end{definition}

\begin{theorem}[\titoloblocchi{Polynomial bound for \LALL}]
 Let $\Pi$ be an \LALL\ proof net of size $s$, level $l$, and depth $d$.
 Then, every canonical reduction is at most $(l+1)s^{{(d+2)}^l}$ steps long.
\end{theorem}
The proof strategy coincides with the one in \cite{BaillotMazza09}, with the following adaptation: the reduction of the garbage collection steps is always delayed till the end.
%
%
%

\adaptedsection{Preliminary notions about \SRN}
\label{section:SRN intermezzo}
We recall from Section~\ref{section:Introduction} that $\SRN^{\numnor{n};\numsafe{s}}$ is the subset of \SRN\ whose terms have normal arity $\numnor{n}$, and safe arity $\numsafe{s}$.
If not otherwise stated, $\vec{t}^{m} = t_1,\ldots, t_m$ will always denote sequences of $m\geq 0$ terms of \SRN.
Moreover, we write $\size{\vec{t}^{m}}\leq l$, for some $l>0$, meaning that the size of every term $t_i$ is not greater than $l$.
Now, from \cite{BellantoniC92}, we recall that, for every $t$ in \SRN${}^{\numnor{n};\numsafe{s}}$, and
$\vec x= x_1,\ldots,x_{\numnor{n}},
 \vec y= y_1,\ldots,y_{\numsafe{s}}$:
{\small
\begin{align}
\label{align:BC-bound}
\size{t(\vec x; \vec y)} \leq
 p_t\left(\size{x_1},\ldots,\size{x_{\numnor{n}}}\right)+\max\left\{\size{y_1},\ldots,\size{y_{\numsafe{s}}}\right\}
\end{align}
}
where $p_t$ is the \emph{characteristic polynomial of $t$} which is non-decreasing and depends on $t$.
We notice that if $u$ is a subterm of $t$, then $\degree{p_u}\leq\degree{p_t}$.
At last, we define the \emph{composition degree} $\degreeC{t}$ and the \emph{recursion degree} $\degreeR{t}$ of $t$, as the functions that count resp. the number of safe composition and recursion schemes inside $t$.
\begin{definition}[\titoloblocchi{The Term Bounding Function $\ob{\cdot}{\cdot}$}]
Let $t$ in \SRN${}^{\numnor{n};\numsafe{s}}$ and $l\geq 0$.
We define $\ob{\cdot}{\cdot}$, that takes $t$ and $l$ as arguments, as
$\ob{t}{l}=p_t(l,\ldots,l)+l$.
\end{definition}
\begin{fact}[\titoloblocchi{$\ob{\cdot}{\cdot}$ Bounds the Output Length of $t\in\SRN$}]
\label{fact:ob bounds the output length}
For every $t$ in \SRN${}^{\numnor{n};\numsafe{s}}$,
$l\geq 0$,
and sequences $\vec{x}, \vec{y}$ such that $\size{\vec x}, \size{\vec y}\leq l$,
we have $\size{t(\vec x; \vec y)} \leq \ob{t}{l}$.
\end{fact}
\begin{definition}[\titoloblocchi{The Net Bounding Function $\nb{\cdot}{\cdot}$}]
\label{definition:The net bounding function nb}
Let $t$ in \SRN${}^{\numnor{n};\numsafe{s}}$ and $l\geq 0$.
We define $\nb{\cdot}{\cdot}$, that takes $t$ and $l$ as arguments, as
$\nb{t}{l}=\ob{t}{\ob{t}{\ldots\ob{t}{l}\ldots}}$,
with $\degreeR{t}+\degreeC{t}$ occurrences of $\ob{t}{\cdot}$.
\end{definition}
\begin{fact}[\titoloblocchi{$\nb{\cdot}{\cdot}$ is a Polynomial}]
\label{fact:nb is a polynomial}
For every fixed $t$ in \SRN${}^{\numnor{n};\numsafe{s}}$, $\nb{t}{l}$ is a polynomial in the free variable $l$, whose degree is $\degree{p_t}^{\degreeC{t}+\degreeR{t}}$.
\end{fact}

\adaptedsection{Preliminary useful \pn s in \LALL}
\label{section:Programming inside tempname}
\complexfigure
{Typed II order affine $\lambda$-terms.}
{fig:linear-lambda-terms}
{\small
\begin{eqnarray}
\nonumber
x^A\in V
       &\Rightarrow& x \in\ALTQ{V}{A}\\
\nonumber
m\geq 1,\
M\in \ALTQ{V}{A},
x_1 \in \ALTQ{V}{A_1},\ldots, x_m\in\ALTQ{V}{A_m}
       &\Rightarrow& (\lambda \otimes_{i=1}^{m}x_i^{A_i}.M) \in \\
\label{align:lambda-terms-abstraction}
       &&\hspace{10mm}\in\ALTQ{V}{A_1\otimes\ldots\otimes A_m\lin A}\\
\nonumber
 M\in \ALTQ{U}{A'\lin A},
  N \in \ALTQ{W}{A'},
  {U\cup W}\subseteq V,
  {U}\cap {W}=\emptyset
  &\Rightarrow& (MN) \in \ALTQ{V}{A}
\\
\nonumber
m\geq 2, \big(1\leq i\neq j\leq m \Rightarrow N_i \in \ALTQ{W_i}{A_i},
\\
\label{align:lambda-terms-tuple}
   W_i\cap W_j=\emptyset,
   W_{i}\subseteq V
  \big)
  &\Rightarrow& (\otimes_{i=1}^{m} N_i)\in \ALTQ{V}{A_1\otimes\ldots\otimes A_m}
\\
\label{align:lambda-generalization}
 M\in\ALTQ{V}{A}
  &\Rightarrow& \Lambda\alpha.M\in\ALTQ{V}{\forall\alpha.A}
\\
\label{align:lambda-extraction}
 M\in\ALTQ{V}{\forall\alpha.A}
  &\Rightarrow& M\{A'\}\in\ALTQ{V}{A\subst{A'}{\alpha}}
\end{eqnarray}
}
We introduce a first set of \pn s useful to define the embedding from \SRN\ to \LALL.
However, whenever neither boxes, nor contractions are used in a given \pn\ $\Pi$, whose conclusion has type $A$, to save space, we represent $\Pi$ by means of a $\lambda$-term. The term belongs to the set $\ALTQ{V}{A}$ of
polymorphic typed \emph{affine} $\lambda$-terms with variables in $V$, patterns, tuples, and type $A\in\FLangQ$.
Figure~\ref{fig:linear-lambda-terms} defines $\ALTQ{V}{A}$.
\eqref{align:lambda-terms-abstraction} introduces $\lambda$-abstractions on a tuple pattern, while \eqref{align:lambda-terms-tuple} introduces tuples.
The application is left-associative. We shall drop useless parenthesis to avoid cluttering the terms.
For any $A$ and $V$, the terms in $\ALTQ{V}{A}$ rewrite under the standard $\beta$-reduction, extended with the following two rules:
(i)
$(\lambda \otimes_{i=1}^{m} x_i.M)(\otimes_{i=1}^{m} N_i)\rightarrow_{\beta} M\substm{N_1}{x_1}{N_m}{x_m}$, where $M\substm{N_1}{x_1}{N_m}{x_m}$ stands for the simultaneous substitution of $N_i$ for $x_i$, with $1\leq i\leq m$,
and
(ii) $(\Lambda\alpha.M)\{B\}\rightarrow_\beta M\subst{B}{\alpha}$.
\capoverso{Booleans.} \label{def:booleans}
The type of booleans is $\B=\forall\gamma.\gamma\lin\gamma\lin\gamma$ whose representative \pn s are:
{\small
\begin{align*}
\FF & = \Lambda\gamma.\lambda x^\gamma y^\gamma.x
           \ \triangleright \ \vdash \B
           & \textrm{(True)}
           \\
\TT & = \Lambda\gamma.\lambda x^\gamma y^\gamma.y
           \ \triangleright \ \vdash \B
           & \textrm{(False)}
           \\
\BoolDiag[b] & = b\,\{\B\otimes\B\}\,(\TT\otimes\TT)\,(\FF\otimes\FF)
           \ \triangleright \B \vdash \B\otimes\B
           & \textrm{(Duplication)}
\end{align*}
}
The \pn\ $\BoolDiag[b]$ duplicates any boolean we may plug into $b$ by a Cut node.
\complexfigure
{(Church) Words.}
{fig:church-bin-num}
{
\hspace{-5mm}
\subfigure[\mbox{$\SW_0[w]$}]{
\label{subfig:SW-0}
  \qquad
  \proofnetpagescaled{0.8}{3.7cm}{
  &\nVoid{out}\\
  &\nRfor{rfor}\\
  &\nRlin{imp1}\\
  &\nRlin{imp1b}\\
  &\nRPA{port1}\\
  &\nRlin{imp2}\\
  &\nAx{ax1}\\
  &\nLlin{llin1}\\
  &&\nAx{ax2}\\
  &&\nLlin{llin2}\\
  &&&\nAx{ax3}\\
  &\nLOC{pax1}&\nLPA{pax2}\\
  &&\nLlin{apply}\\
  \nWea{weak}&&\nLfor{lfor}&\nAx{ax4}\\
  &\nCon{con}\\
  &&\nVoid{input}
  \ncline{rfor}{out} \nbput{\NC}
  \ncline{imp1}{rfor}
  \ncline{imp1b}{imp1}
  \ncline{port1}{imp1b} \nbput{\pa{(\alpha\lin\alpha)}}
  \ncline[offset=2pt]{con}{pax1}  \naput{\oc(\alpha\lin\alpha)}
  \ncangle[offsetA=-2pt,angleA=90,angleB=-90,armB=3mm]{con}{ax4}
  \ncangle[angleA=90,angleB=0]{ax4}{apply}
  \ncldownup[armA=9mm]{imp1}{con} \nbput[npos=0.5]{\oc{(\alpha\lin\alpha)}}
  \ncldownup[armA=8mm]{imp1b}{weak} \nbput[npos=0.5]{\oc{(\alpha\lin\alpha)}}
  \ncline{apply}{pax2} \nbput{\pa(\alpha\lin\alpha)}
  \ncline{lfor}{apply} \ncput[npos=0.4]{\oc{(\alpha\lin\alpha)}\lin\pa{(\alpha\lin\alpha)}}
  \ncline{input}{lfor} \nbput\NC
  \ncline{imp2}{port1} \nbput{(\alpha\lin\alpha)}
  \ncline{ax1}{imp2} \nbput{\alpha}
  \ncline{llin1}{ax1} \nbput{\alpha}
  \ncline{pax1}{llin1} \naput{\alpha\lin\alpha}
  \ncangle[angleA=90,angleB=0]{ax2}{llin1} \nbput[npos=0.5]{\alpha}
  \ncline{llin2}{ax2} \nbput{\alpha}
  \ncline{pax2}{llin2} \naput{\alpha\lin\alpha}
  \ncangle[angleA=90,angleB=0]{ax3}{llin2} \nbput[npos=0.5]{\alpha}
  \ncldownup[armB=2.5mm]{imp2}{ax3} \nbput[npos=3.5]{\alpha}
  \ncangle[linewidth=1.5pt,linestyle=dotted,angleA=180,angleB=180,arm=8mm]{-}{port1}{pax1}
  \ncline[linewidth=1.5pt,linestyle=dotted]{-}{pax1}{pax2}
  \ncangle[linewidth=1.5pt,linestyle=dotted,angleA=0,angleB=0,arm=25mm]{-}{pax2}{port1}
  }
 } 
\subfigure[\mbox{$\RevW[w]$}]{
\label{subfig:RevW}
 \proofnetpagescaled[colsep=1cm]{0.8}{3.8cm}{
  \nVoid{o11}\\
  \nRfor{rf31}\\
  \nRlin{rl41}\\
  \nRlin{rl51}\\
  \nRPA{rp61  }\\
  \nLlin{ll71 }&\simplenet{id72}{\lambda x.x}\\
  \nLPA{lp81  }\\
  \nLlin{ll91 }&\simplenet[11mm]{st92 }{\texttt{Step}_1}\\
  \nLlin{ll101}\\
  \nLfor{lf111}&\simplenet[11mm]{st112}{\texttt{Step}_0}\\
  \\ \\
  \nVoid{i121}
  \\ \\
  \ncline{rf31 }{o11 }
    \nbput{\NC}
  \ncline{rl41 }{rf31 }
  \ncline{rl51 }{rl41 }
  \ncline{rp61 }{rl51 } \nbput{\S(\alpha\lin\alpha)}
  \ncline{ll71 }{rp61 }
  \ncline{lp81 }{ll71 } \nbput{\alpha\lin\alpha}
  \ncline{ll91 }{lp81 }
  \ncline{ll101}{ll91 }
  \ncline{lf111}{ll101} \nbput{\NC\{\alpha\lin\alpha\}}
  \ncline{i121}{lf111}
    \nbput{\NC}
  \ncangle[angleA=90,angleB=45]{id72 }{ll71 } \nbput[npos=1.3]{(\alpha\lin\alpha)\lin(\alpha\lin\alpha)}
  \ncangle[angleA=90,angleB=45]{st92 }{ll91 }
  \ncangle[angleA=90,angleB=45]{st112}{ll101}
  \ncangles[angleA=180,angleB=-90,armA=4mm]{rl41}{st112}
  \ncloop[angleA=180,angleB=0,armA=5mm,armB=2mm,loopsize=64mm]{rl51}{st92 }
    \ncangle[linewidth=1.5pt,linestyle=dotted,angleA=180,angleB=180]{-}{rp61}{lp81}
    \ncangle[linewidth=1.5pt,linestyle=dotted,angleA=0,angleB=0,armB=27mm]{-}{lp81}{rp61}
  }
 } 
\begin{tabular}{c}
\subfigure[$\ZW$]{
\label{subfig:ZW}
 \proofnetpagescaled[colsep=1cm]{0.8}{2.5cm}{
  &&\nVoid{out}\\
  &&\nRfor{rfor}\\
  &&\nRlin{lambda1}\\
  &&\nRlin{lambda2}\\
  &&\nRPA{port}\\
  &&\nRlin{lambda3}\\
  \nWea{weak1} &\nWea{weak2} &\nAx{ax}\\
  \ncline{rfor}{out} \nbput{\NC}
  \ncline{lambda1}{rfor}
  \ncline{lambda2}{lambda1}
  \ncline{port}{lambda2}
    \naput{\pa{(\alpha\lin\alpha)}}
  \ncline{lambda3}{port}
  \ncline{ax}{lambda3} \nbput{\alpha}
  \ncldownup[armA=25mm]{lambda1}{weak1} \nbput[npos=0.5]{\oc{(\alpha\lin\alpha)}}
  \ncldownup[armA=15mm]{lambda2}{weak2} \nbput[npos=0.5]{\oc{(\alpha\lin\alpha)}}
  \ncldownup[armA=1mm]{lambda3}{ax} \nbput[npos=3.5]{\alpha}
  \ncloop[linewidth=1.5pt,linestyle=dotted,loopsize=25mm,angleA=180,angleB=0]{-}{port}{port}
  }
 } 
\\
\subfigure[\texttt{Step}${}_i$ with $i\in\{0,1\}$]{
\label{subfig:Step-i}
 \proofnetpagescaled[colsep=1cm]{0.8}{3.5cm}{
  \nVoid{o11}\\
  \nROC{ro21  }\\
  \nRlin{rl31 }\\
  \nRlin{rl41 }\\
  \nLlin{ll51 }&\nLlin{ll52 }&\nAx{ax53 }\\ \\
               &\nLOC{lo62  }\\
               &\nVoid{i72}
  \ncline{ro21 }{o11 }
    \nbput{\ocs((\alpha\lin\alpha)\lin \alpha\lin\alpha)}
  \ncline{rl31 }{ro21 }
    \nbput{(\alpha\lin\alpha)\lin \alpha\lin\alpha}
  \ncline{rl41 }{rl31 }
  \ncline{ll51 }{rl41 }
  \ncangle[angleA=90,angleB=45]{ll52 }{ll51 }
  \ncangle[angleA=90,angleB=45]{ax53 }{ll52 }
  \ncline{lo62}{ll52}
  \ncline{i72}{lo62}
    \nbput{\ocs(\alpha\lin\alpha)}
  \ncangles[angleA=180,angleB=-90,armA=4mm]{rl31}{ll51}
  \ncangles[angleA=180,angleB=-90,armA=5mm,armB=6mm]{rl41}{ax53 }
    \ncangle[linewidth=1.5pt,angleA=0,angleB=0,armB=18mm]{-}{ro21}{lo62}
    \ncangle[linewidth=1.5pt,angleA=180,angleB=180,armB=10mm]{-}{lo62}{ro21}
  }
 } 
\end{tabular}
} 

\capoverso{Church words or, simply, words.}
The type of words is
$\NC=\forall\alpha.\oc(\alpha\multimap\alpha)\multimap\oc(\alpha\multimap\alpha)\multimap\pa(\alpha\multimap\alpha)$.
\condinc{
Figure~\ref{fig:church-bin-num} introduces the word $\ZW$ and the successors $\SW_0^w$, $\SW_1^w$.
}
{
}
Figure~\ref{subfig:SW-0} introduces the successor $\SW_0[w]$, where $w$ identifies the lowermost dangling edge.
It should be trivial to recover $\SW_1[w]$ from $\SW_0[w]$.
Figure~\ref{subfig:ZW} introduces $\ZW$.
If $w$ is a natural number in binary notation, $\ncof w$ is its usual representation by a \pn.
Figures~\ref{subfig:RevW} and~\ref{subfig:Step-i} introduce \pn s that invert the bits inside any $\ncof w$, plugged by Cut into the dangling input of $\RevW[w]$.
\capoverso{Scott words.}
Intuitively, the type $\ND$ of Scott words describes a tuple of booleans.
On Scott words we have the following \pn s:
{\small
\begin{align*}
\ZD & = \Lambda\alpha.\lambda x^\alpha y^{\B\lin\ND\lin\alpha}.x
        \ \triangleright \ \vdash \ND
    & \textrm{(Empty Scott word)}
\\
\SD_0[s] & = \Lambda\alpha.\lambda x^{\alpha}y^{\B\lin\ND\lin\alpha}.y\,\texttt{F}\,s
            \, \triangleright \ND \vdash \ND
    & \textrm{(Successor zero)}
\\
\SD_1[s] & = \Lambda\alpha.\lambda x^{\alpha}y^{\B\lin\ND\lin\alpha}.y\,\texttt{T}\,s
            \, \triangleright \ND \vdash \ND
    & \textrm{(Successor one)}
\\
\PD[s]   & = s\,\{\ND\}\, \ZD\,(\lambda b^\B w^\ND.w)
            \, \triangleright \ND \vdash \ND
    & \textrm{(Predecessor)}
\\
\BD[s,x,y] & = \PPD[s]\,\{\ND\}\, \ZD\,(\lambda b^\B .b\,\{\ND\}\,y\,x)
            \, \triangleright \ND,\ND,\ND \vdash \ND
    & \textrm{(Conditional)}
\\
\PPD[s] & = s\,\{\ND\}\, \SD_0[\ZD]\,\big(\lambda b^\B\lambda w^\ND.b\{\ND\lin\ND\}\\
&\hspace{3cm}
 (\lambda x^\ND.\SD_0[x])\,(\lambda x^\ND.\SD_1[x])\, w\big)
  \, \triangleright \ND \vdash \ND
    & \textrm{(Preprocessing)}
\end{align*}
}
We remark that $\SD_0[s]$ adds to $s$ the least significant bit $\TT$, which stands for the digit $0$,
and $\SD_1[s]$ adds $\FF$, instead, which stands for $1$.
$\PD[s]$ shifts $s$ to its right deleting the \emph{least significant} bit. So:
\begin{remark}\label{remark:Bits are reversed}
 A Scott word is in fact a stack of bits, the least significant bit being on the top of the stack.
\end{remark}
Moreover, $\BD[s,x,y]$ branches a computation, depending on the value of $s$.
It yields $x$ if the least significant bit of $s$ is $0$, \emph{or} if $s=\ZD$,
while it yields $y$ if the least significant bit of $s$ is $1$.
The preprocessing avoids to return $\ZD$: if $s=\ZD$, then $s$ becomes $\SD_0[\ZD]$.
Also, the three assumptions of type $\ND$ in $\ND,\ND,\ND \vdash \ND$ specify the type of $s, x$, and $y$, respectively.
\begin{fact}[\titoloblocchi{Relation between naturals, Scott words, and words-as-terms}]
 \label{fact:SRN-numerals}
 Every sequence $(d_1,\ldots,d_l)$ with $d_1,\ldots, d_{l}\in\{0,1\}$ and $l\geq 0$, identifies uniquely a number
 $n = 2^{l-1}\cdot d_{l}+\cdots+2^{0}\cdot d_{1} \in\N$.
 So, both the term of \SRN\ $\Sucg{d_{l}}(;\ldots\Sucg{d_{1}}(;\Zero)\ldots)$ and
 the Scott word $\nsof{n}$ identify $n$, too.
 We say that the sequence, as well as the Scott number and the \SRN\ term, \emph{represent} $n$.
\end{fact}
We underline that an infinite number of sequences, and of terms, represent the same $n$.

\capoverso{Scott words to words.} \label{def:DtoW}
For any $l\geq 0$, $\DtoW_{l}[s] \triangleright \, \ND \vdash \NC$ is inductively defined on $l$:
{
\small
\begin{align*}
\DtoW_0[s] & = \ZW
\\
\DtoW_{l}[s] & =
s\,\{\NC\}\,\ZW\,(\lambda x^\B y^\ND.x\,(\lambda z^{\NC}.\SW_0[z])
                                                   \,(\lambda z^{\NC}.\SW_1[z])
                                                   \, \DtoW_{l-1}[s])
\end{align*}
}
The \pn\ $\DtoW_l[s]$ normalizes to the word $\ncof w$, for any Scott word \emph{at most as long as} $l$.
\capoverso{Duplicating Scott words.} \label{def:DD}
For any $l\geq 0$, the \pn\ $\DD_l[s]$ is inductively defined on $l$:
{
\small
\begin{align*}
\DD_0[s]   & = \ZD\otimes\ZD\\
\DD_{l}[s] & = s\, \{\ND\otimes\ND\}\,(\ZD\otimes\ZD)\,
          \big(\lambda b^\B s^\ND.
           b\{\ND^2\lin\ND^2\}\, (\lambda x^\ND\otimes y^\ND. \SD_0[x]\otimes\SD_0[y])
        \\
          &\phantom{
                     s\, [\ND\otimes\ND]\,(\ZD\otimes\ZD)(\lambda b^\B s^\ND.b[\ND^2\lin\ND^2]\,
                     \quad \ \,
                   }
          (\lambda x^\ND\otimes y^\ND. \SD_1[x]\otimes\SD_1[y])
          \,\DD_{l-1}[s]
          \big)
\end{align*}
}
such that $\DD_{l}[s]\triangleright \, \ND \vdash \ND^2$, where $\ND^2 = \ND\otimes \ND$.
The \pn\ $\DD_l[s]$ builds two copies of any Scott word \emph{at most as long as} $l$.
The generalization $\DDg{k}_{l}[s]\triangleright \, \S^{k}\ND \vdash \S^{k}\ND\otimes\S^{k}\ND$ of $\DD_{l}[s]$ duplicates a given Scott word \emph{at most as long as} $l$ which lies inside $k\geq 0$ paragraph boxes. Specifically,
$\DDg{0}_{l}[s]$ is $\DD_{l}[s]$, while $\DDg{k}_{l}[s]$ is in Figure~\ref{fig:safe-contraction},
with $k>0$, which is the only \pn\ that exploits the \emph{fuzzy borders} of paragraph boxes.
By induction on $k$,
$\size{\DDg{k}_l[s]}=19+89l+3k \in O(l)$.
\complexfigure
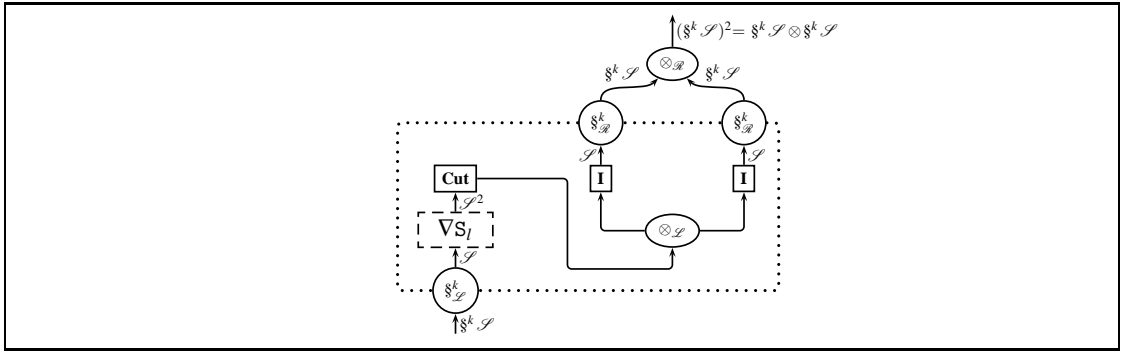
{The generalized duplication $\DDg{k}_{l}[s]$ of Scott words.}
{fig:safe-contraction}
{
\begin{center}
  \proofnetpagescaled{0.8}{4cm}{
                         &&&&               &\nVoid{out}\\
                         &&&&               &\nRtens{rte}\\
                         &&&&\nRPAm{rpa1}{k}&            &\nRPAm{rpa2}{k}\\
    \nCut{cut}           &&&&\nAx{a1}       &            &\nAx{a2}\\
    \simplenet{dd}{\DD_l}&&&&               &\nLtens{lte2}\\
    \nLPAm{leftpa}{k}\\
    \nVoid{input}
    \ncline{rte}{out}
        \nbput{(\pas^k\ND)^2 = \ \pas^k\ND\, \otimes\, \pas^k\ND}
    \ncupl{rpa1}{rte}  \naput{\pas^k\ND}
    \ncupr{rpa2}{rte}  \nbput{\pas^k\ND}
    \ncline{a1}{rpa1}  \naput{\ND}
    \ncline{a2}{rpa2}  \nbput{\ND}
    \ncangle[angleA=180,angleB=-90]{lte2}{a1}
    \ncangle[angleA=0,angleB=-90]{lte2}{a2}
    \ncangles[angleA=0,angleB=-90,armA=15mm]{cut}{lte2}
    \ncline{dd}{cut} \nbput{\ND^2}
    \ncline{leftpa}{dd}  \nbput{\ND}
    \ncline{input}{leftpa}  \nbput{\pas^k\ND}
    \ncline[linewidth=1.5pt,linestyle=dotted]{-}{rpa1}{rpa2}
    \ncangle[linewidth=1.5pt,linestyle=dotted,angleA=180,angleB=180,armB=30mm]{-}{leftpa}{rpa1}
    \ncangle[linewidth=1.5pt,linestyle=dotted,angleA=0,angleB=0,armB=2mm]{-}{leftpa}{rpa2}
  }
\end{center}
}

\capoverso{Coercing Scott words.}
For any $k, l\geq 0$, we define $\CoerceD^{k}_l[s]\triangleright\,\ND\vdash\S^k\ND$ by cases on $k$, and by induction on $l$.
If $k=0$, then $\CoerceD^{0}_l[s]$ is the node Axiom.
Otherwise, the \pn\ is in Figure~\ref{fig:scott-words-coercion}, where,
for $i\in\{0,1\}$,
$\lambda s.\S^k(\SD_i[s])\,\triangleright\,\vdash \S^k\ND\lin\S^k\ND$ \emph{denotes} the \pn\ that we build by:
(i) enclosing $\SD_i[s]$ into $k$ paragraph boxes to get $\S^k(\SD_i[s])\,\triangleright\,\S^k\ND\vdash\S^k\ND$, and
(ii) adding an Implication R to $\S^k(\SD_i[s])$ so to close it and get its type $\S^k\ND\lin\S^k\ND$.
The \pn\ $\CoerceD^{k}_l[s]$ reconstructs a given Scott word \emph{at most as long as} $l$ into an identical Scott word inside $k$ paragraph boxes.
We can show that $\size{\CoerceD^{k}_l[s]}=43l+3kl \in O(l)$.
\complexfigure
{The coerce \pn\ $\CoerceD^k_l[s]$ on Scott words.}
{fig:scott-words-coercion}
 {
  \centering
  \proofnetpagescaled{0.8}{.7\textwidth}{
    \nVoid{o11}\\
    \nLlin{il21}&                 &&\nRlin{ir23}\\
                &                 &&\nRlin{ir33}\\
                &                 &&\nLlin{il43}\\
                &                 &&\nLlin{il53}\\
    \nLlin{il61}&\nRPAm{parm62}{k}&&\nLlin{il63}\\
    \nLfor{fl71}&\simplenet[20mm]{net72}{\ZD}
                                  &&\nLfor{fl73}&\simplenet[25mm]{net74}{\lambda s.\S^k(\SD_0[s])}
                                                      &\simplenet[25mm]{net75}{\lambda s.\S^k(\SD_1[s])}
                                                            &\simplenet[18mm]{net76}{\CoerceD^k_{l-1}[s]}\\
    \nVoid{i81}
    \ncline{il21}{o11}
      \naput{\pas^k\ND}
    \ncline{il61}{il21}
      \naput{(\B\lin\ND\lin\pas^k\ND)\lin\pas^k\ND}
    \ncline{fl71}{il61}
      \naput{\ND\{\pas^k\ND\}}
    \ncline{i81}{fl71}
      \naput{\ND}
    \ncangle[angleA=90,angleB=45,armA=2mm]{parm62}{il61}
      \nbput{\pas^k\ND}
    \ncline{parm62}{net72}
    \ncangle[angleA=90,angleB=45,armA=3mm]{ir23}{il21}
      \nbput{\B\lin\ND\lin\pas^k\ND}
    \ncline{ir33}{ir23}
    \ncline{il43}{ir33} \nbput{\pas^k\ND}
    \ncline{il53}{il43}
    \ncline{il63}{il53}
    \ncline{fl73}{il63}
    \ncangles[angleA=180,angleB=-90,armA=2mm]{ir23}{fl73}
      \nbput[npos=3.5]{\B}
    \ncangles[angleA=180,angleB=-90,armA=1mm,armB=4mm]{ir33}{net76}
    \ncangle[angleA=90,angleB=0]{net74}{il63}
      \nbput{\pas^k\ND\lin\pas^k\ND}
    \ncangle[angleA=90,angleB=0]{net75}{il53}
      \nbput{\pas^k\ND\lin\pas^k\ND}
    \ncangle[angleA=90,angleB=0]{net76}{il43}
      \nbput{\pas^k\ND}
    \ncloop[linewidth=1.5pt,linestyle=dotted,loopsize=15mm,angleA=180,angleB=0,arm=9mm]{-}{parm62}{parm62}
  }
}
\capoverso{Lifting.}
Let $\Pi\, \triangleright\, {\vec \ND}^\numnor{n}, {\vec{ \S^k\ND}}^\numsafe{s}\vdash \S^k\ND$ for some $\numnor{n}, \numsafe{s}, k\geq 0$.
For every $k'\geq 0$ we can build
$\Lift{k}{k'}[\Pi]\, \triangleright\, {\vec \ND}^\numnor{n}, {\vec{ \S^{k+k'}\ND}}^\numsafe{s}\vdash \S^{k+k'}\ND$ by:
(i) enclosing $\Pi$ into $k'$ paragraph boxes, getting $\Pi'$, and
(ii) plugging the conclusion of $\CoerceD^{k'}_{l}[s]$, using Cut, into every of the $\numnor{n}$ premises with type $\S^{k'}\ND$ of $\Pi'$.
The \pn\ $\Lift{k}{k'}[\Pi]$ is $\Pi$ deepened inside $k'$ paragraph boxes.
The final \pn\ is in Figure~\ref{fig:lifting}.
\complexfigure
{The \emph{Lifting} $\Lift{k}{k'}{[\Pi]}$ of a proof net $\Pi$.}
{fig:lifting}
{
  \begin{center}
  \proofnetpagescaled{0.8}{8cm}{
    &&&\nVoid{out}\\
    &&&\nRPAm{po1}{k'}\\
    &&&\simplenet{pi}{\Pi}\\ \\
    \nLPAm{p1}{k'} & &\nLPAm{p2}{k'} &&\nLPAm{p3}{k'} &&\nLPAm{p4}{k'}\\
    \simplenet[1.6cm]{coe1}{\CoerceD^{k'}_{l}}&&\simplenet[1.6cm]{coe2}{\CoerceD^{k'}_{l}}\\
    \nVoid{i1} & \ldots &\nVoid{i2} &&\nVoid{i3} &\ldots &\nVoid{i4}
    \ncline{pi}{po1} \nbput{\S^{k}\ND}
    \ncup[offsetB=.3,armB=2mm]{p1}{pi} \naput{\ND}
    \ncup[offsetB=.1,armB=3mm]{p2}{pi}  \nbput{\ND}
    \ncup[offsetB=-.1,armB=3mm]{p3}{pi} \naput{\S^{k}\ND}
    \ncup[offsetB=-.3,armB=2mm]{p4}{pi} \nbput{\S^{k}\ND}
    \ncline{po1}{out} \nbput{\S^{k+k'}\ND}
    \ncline{coe1}{p1} \naput{\S^{k'}\ND}
    \ncline{coe2}{p2} \naput{\S^{k'}\ND}
    \ncline{i1}{coe1} \naput{\ND}
    \ncline{i2}{coe2} \naput{\ND}
    \ncline{i3}{p3} \nbput{\S^{k+k'}\ND}
    \ncline{i4}{p4} \nbput{\S^{k+k'}\ND}
    \ncangle[linewidth=1.5pt,linestyle=dotted,angleA=180,angleB=180]{-}{po1}{p1}
    \ncline[linewidth=1.5pt,linestyle=dotted]{-}{p1}{p2}
    \ncline[linewidth=1.5pt,linestyle=dotted]{-}{p2}{p3}
    \ncline[linewidth=1.5pt,linestyle=dotted]{-}{p3}{p4}
    \ncangle[linewidth=1.5pt,linestyle=dotted,angleA=0,angleB=0,armB=36mm]{-}{p4}{po1}
  }
  \end{center}
}

\par
Notice that $\size{\Lift{k}{k'}[\Pi]}=|\Pi|+k'(\numnor{n}+\numsafe{s}+1)+\numnor{n}\size{\CoerceD^{k'}_{l}[s]} \in O\left(\size{\Pi}+l\right)$.
\capoverso{Contracting the premises of a \pn.}
Let $\Pi\, \triangleright\, {\vec A}, \S^k\ND, \S^k\ND, {\vec A'} \vdash A$ for some $l, k\geq 0$.
We can build $\CD^k_l[\Pi]\, \triangleright\, {\vec A}, \S^k\ND, {\vec A'} \vdash A$ by:
(i)  writing $\Pi'$ which is $\Pi$ with a new Tensor L between the two outlined premises of type $\S^k\ND$, and
(ii) plugging the conclusion of $\DDg{k}_{l}[s] \, \triangleright \, \S^k\ND\vdash \S^k\ND\otimes \S^k\ND$, by a Cut, into the premise of the new Tensor L in $\Pi'$.
Notice that $\size{\CD^k_l[\Pi]}=|\Pi|+ \size{\DDg{k}_{l}[s]} + 2 \in O(l)$.
\adaptedsection{\texorpdfstring{The embedding $\srntolallg$ from \SRN\ to \LALL}
                               {The embedding from \SRN\ to \LALL}}
\label{section:Defining the embedding srntolall from SRN to LALL}
The goal is to compositionally embed \SRN\ into \LALL, with a map as much analogous as possible to the natural, and inductively defined one from \BCminus\ into \ILAL\ \cite{MurawskiOng00}.
For any fixed $\numnor{n}$ and $\numsafe{s}$, the map $\srntolallg$ takes a term $t$ of \SRN${}^{\numnor{n};\numsafe{s}}$ as first and $l\geq 0$ as second argument, and yields a \pn\
$\srntolall{t}{l}\,\triangleright \,{\vec \ND}^\numnor{n},{\vec{\S^{k}\ND}}^\numsafe{s}\vdash \S^{k}\ND$, for some $k$.
We define the map inductively on the first argument.
\capoverso{The base cases of $\srntolall{\cdot}{\cdot}$.}
Some of them are straightforward:
{\small
\begin{center}
 \begin{tabular}{ll}
$\srntolall{\Zero}{l} = \ZD \ \triangleright\,\vdash\ND$
&
$\srntolall{\Suci}{l} = \SD_i[s] \,\triangleright\, \ND\vdash\ND \qquad \qquad (i\in\{0,1\})$
\\
$\srntolall{\Pred}{l} = \PD[s] \,\triangleright\, \ND\vdash\ND$
&
$\srntolall{\Bran}{l} = \BD[s,x,y] \,\triangleright\, \ND,\ND,\ND \vdash\ND$
\end{tabular}
\end{center}
}
\noindent
where, $s,x,y$ denote the inputs of the \pn s they appear into.
Concerning the projection, $\srntolall{\Proj{i}{\numnor{n};\numsafe{s}}}{l}$ is an Axiom that connects the $i$-th input to the conclusion, erasing all the other inputs by Weakening.
An example with $1\leq i\leq \numnor{n}$, and, notice, $k=0$ is:
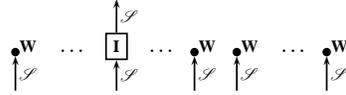
{
 \begin{center}
  \proofnetpagescaled{0.8}{\textwidth}{
                 &           &\nVoid{o13}\\
    \nWea{we21}{}&\quad\ldots&\nAx{ax23} &\ldots&\nWea{we25}{}&&\nWea{we27}{}&\quad\ldots&\nWea{we29}{}\\
    \nVoid{i31}  &           &\nVoid{i33}&      &\nVoid{i35}  &&\nVoid{i37}  &           &\nVoid{i39}
    \ncline{ax23}{o13}
      \nbput{\ND}
    \ncline{i33}{ax23}
      \nbput{\ND}
    \ncline{i31}{we21}
      \nbput{\ND}
    \ncline{i35}{we25}
      \nbput{\ND}
    \ncline{i37}{we27}
      \nbput{\ND}
    \ncline{i39}{we29}
      \nbput{\ND}
 }
 \end{center}
}

\capoverso{The case of $\srntolallg$ on the composition.}
We now focus on $t = \Comp{t',u_1,\ldots,u_{\numnor{m}},v_1,\ldots,v_{\numsafe{r}}}$ such that $t'$ be in \SRN${}^{\numnor{m};\numsafe{r}}$.
Without loss of generality, we show how to build $\srntolall{t}{l}$ by assuming
$\numnor{m}=\numsafe{n}=\numnor{s}=1$, and $\numsafe{r}=2$.
By induction we have:
{\small
$$
\begin{array}{ccc}
\srntolall{t'}{\ob{t}{l}} \triangleright \,\ND, \S^{k'}\ND, \S^{k'}\ND \vdash \S^{k'}\ND
& \quad &
\srntolall{u_1}{l} \triangleright \,\ND \vdash \S^{k_u}\ND
\\
\srntolall{v_i}{l} \triangleright \,\ND, \S^{k_i}\ND \vdash \S^{k_i}\ND
&& (i\in\{1, 2\})
\end{array}
$$
}
By letting $k=\max\{k', k_u, k_1, k_2\}$, we get:
{\small
\begin{center}
\begin{tabular}{ll}
$\Lift{k' }{(k-k')}[\srntolall{t' }{\ob{t}{l}}] \triangleright \,\ND, \S^{k}\ND, \S^{k}\ND \vdash \S^{k}\ND$
&
$\Lift{k_u}{(k-k_u)}[\srntolall{u_1}{l}] \triangleright \,\ND \vdash \S^{k}\ND$
\\
$\Lift{k_i}{(k-k_i)}[\srntolall{v_i}{l}] \triangleright \,\ND, \S^{k}\ND \vdash \S^{k}\ND$
& $(i\in\{1, 2\})$
\end{tabular}
\end{center}
}
Next, if we build $\Pi'$ in Figure~\ref{fig:translating-safe-composition}, then
$\srntolall{t}{l}$ is $\CD^{2k}_l[\CD^{0}_l[\CD^{0}_l[\Pi']]]$.
The two occurrences of $\CD^{0}_l$ contract three ``normal'' inputs. One is from $\srntolall{u_1}{l}$. The other two are from $\srntolall{v_1}{l}, \srntolall{v_2}{l}$.
The occurrence of $\CD^{2k}_l$ contracts the single ``safe'' input of $\srntolall{v_1}{l}$ and $\srntolall{v_2}{l}$.
\complexfigure
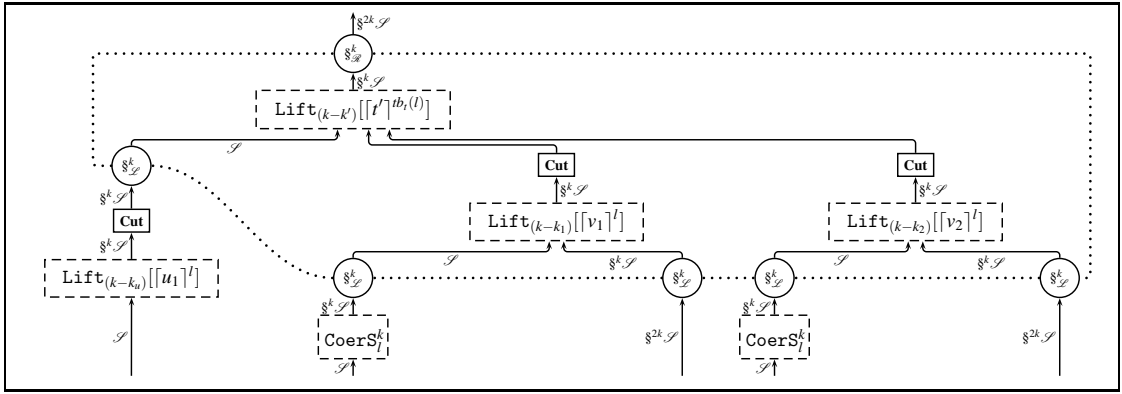
{The (partial) translation of $\Comp{t',u_1,v_1,v_{2}}$ with missing contractions.}
{fig:translating-safe-composition}
{
 \begin{center}
  \proofnetpagescaled{0.69}{\textwidth}{
                 &&\nVoid{out}\\
                 &&\nRPAm{po1}{k}\\
                 &&\simplenet[35mm]{pi}{\Lift{k' }{(k-k')}[\srntolall{t' }{\ob{t}{l}}]}\\
    \nLPAm{p1}{k}&&&\nCut{cut3}                                   &&          &&\nCut{cut4}\\
    \nCut{cut2}  &&&\simplenet[31mm]{tt}{\Lift{k_1}{(k-k_1)}[\srntolall{v_1}{l}]}
                                                                  &&          &&\simplenet[31mm]{tt1}{\Lift{k_2}{(k-k_2)}[\srntolall{v_2}{l}]}\\
    \simplenet[31mm]{ss}{\Lift{k_u}{(k-k_u)}[\srntolall{u_1}{l}]}
                 &&\nLPAm{pb7}{k} &&\nLPAm{pb8}{k}&&\nLPAm{pb9}{k}&&          &\nLPAm{pb10}{k}\\
                 &&\simplenet[11mm]{coe2}{\CoerceD^{k}_{l}}       &&          &&\simplenet[11mm]{coe3}{\CoerceD^{k}_{l}}\\
    \nVoid{i1}   &&\nVoid{i9}&&\nVoid{i4}&                        &\nVoid{i5} &&&\nVoid{i7}
    \ncline{po1}{out} \nbput{\S^{2k}\ND}
    \ncline{pi}{po1} \nbput{\S^{k}\ND}
    \ncup[offsetB=-.3,armB=3mm]{cut3}{pi}
    \ncup[offsetB=-.7,armB=2mm]{cut4}{pi}
    \ncangle[linewidth=1.5pt,linestyle=dotted,angleA=180,angleB=180]{-}{po1}{p1}
    \nccurve[linewidth=1.5pt,linestyle=dotted,angleA=0,angleB=180]{-}{p1}{pb7}
    \ncline[linewidth=1.5pt,linestyle=dotted]{-}{pb7}{pb8}
    \ncline[linewidth=1.5pt,linestyle=dotted]{-}{pb8}{pb9}
    \ncline[linewidth=1.5pt,linestyle=dotted]{-}{pb9}{pb10}
    \ncangle[linewidth=1.5pt,linestyle=dotted,angleA=0,angleB=0,armB=2mm]{-}{po1}{pb10}
    \ncup[offsetB=.3,armB=2mm]{p1}{pi} \nbput{\ND}
    \ncline{cut2}{p1} \naput{\S^k\ND}
    \ncline{ss}{cut2} \naput{\S^k\ND}
    \ncline{i1}{ss} \naput\ND
    \ncline{tt}{cut3} \nbput{\S^{k}\ND}
    \ncup[armB=2mm,offsetB=4pt]{pb7}{tt} \nbput\ND
    \ncup[armB=2mm,offsetB=-4pt]{pb8}{tt} \naput[npos=1.5]{\S^k\ND}
    \ncline{coe2}{pb7} \naput{\S^k\ND}
    \ncline{i4}{pb8}  \naput{\S^{2k}\ND}
    \ncline{i9}{coe2}  \naput{\ND}
    \ncline{tt1}{cut4} \nbput{\S^{k}\ND}
    \ncup[armB=2mm,offsetB=4pt]{pb9}{tt1} \nbput\ND
    \ncup[armB=2mm,offsetB=-4pt]{pb10}{tt1} \naput[npos=1.5]{\S^k\ND}
    \ncline{coe3}{pb9} \naput{\S^k\ND}
    \ncline{i5}{coe3} \naput{\ND}
    \ncline{i7}{pb10} \naput{\S^{2k}\ND}
  }
 \end{center}
}

We insist remarking the existence of $\srntolall{t}{l}$ for any $\numnor{m},\numnor{n},\numsafe{r},\numsafe{s}$.
One can count:
$\size{\srntolall{t}{l}} \leq
\size{\Lift{k' }{(k-k')}[\srntolall{t' }{\ob{t}{l}}]}
+\sum_i^{\numnor{m}}\size{\Lift{k_i}{(k-k_i)}[\srntolall{u_i}{l}]}
+\sum_j^{\numsafe{r}}\size{\Lift{k_j}{(k-k_j)}[\srntolall{u_j}{l}]}
+k(1+\numnor{n}+\numsafe{s'}\numnor{n}+\numsafe{s'}\numsafe{s}) 
+\numsafe{s'}\numnor{n}\size{\CoerceD^{k}_{l}[s]}
$.
So, it follows
$\size{\srntolall{t}{l}} \in
O\left(
		\size{\srntolall{t' }{\ob{t}{l}}}
		+\sum_i^{\numnor{m}}\size{\srntolall{u_i}{l}}
		+\sum_j^{\numsafe{r}}\size{\srntolall{v_j}{l}}
		+\ob{t}{l}+l
		\right)$.
\capoverso{The case of $\srntolallg$ on the recursion.}
Let $t = \Rec{u_{\varepsilon},u_0,u_1}$ with $u_{\varepsilon}\in \SRN^{\numnor{n};\numsafe{s}}, u_0, u_1\in \SRN^{\numnor{n}+1;\numsafe{s}+1}$.
As for composition, we set $\numnor{n}=\numsafe{s}=1$ which is general enough to show the key point of the embedding.
In the course of the iteration unfolding that $\srntolall{t}{l}$ carries out, the safe argument gets duplicated, so we must contract it.
By induction:
{\small
\begin{center}
\begin{tabular}{cccc}
$\srntolall{u_\varepsilon}{l} \triangleright \,\ND, \S^{k_{\varepsilon}}\ND \vdash \S^{k_{\varepsilon}}\ND$
&&&
$\srntolall{u_i}{\ob{t}{l}} \triangleright \,\ND,\ND, \S^{k_{i}}\ND, \S^{k_{i}}\ND \vdash \S^{k_{i}}\ND$
                   \qquad $(i\in\{0, 1\})$
\end{tabular}
\end{center}
}
By letting $k=\max\{k_{\varepsilon}, k_{0}, k_{1}\}$, and using $\Lift{\cdot}{\cdot}[\cdot]$ in analogy to the translation of the composition,
$\srntolall{t}{l}$ is in Figure~\ref{fig:translating-safe-recursion}.
The Scott word that drives the recursion unfolding, becomes a word, and, then, it is necessary to reverse it by $\RevW[w]$. Otherwise we would unfold the iteration according to a wrong bit order, as implied by Remark~\ref{remark:Bits are reversed}.
Moreover,
(i) $\Pi$ projects the rightmost $\numnor{n}+\numsafe{s}+1$-th element of type $A$ it gets in input and which contains the result, and
(ii) the two \pn s $\nRMtens{p}, \nLMtens{q}$ are obvious trees of Tensor R and L nodes.
Finally, we can prove
$\size{\srntolall{t}{l}} \in
O\left(
      \size{\srntolall{u_\varepsilon}{l}}
      +\size{\srntolall{u_0}{\ob{t}{l}}}
      +\size{\srntolall{u_1}{\ob{t}{l}}}
      +\ob{t}{l}
      \right)$.
\complexfigure
{Safe recursion.}
{fig:translating-safe-recursion}
{
\hspace{-4mm}
\begin{tabular}{ccc}
 \begin{minipage}{.6\textwidth}
 {
  \proofnetpagescaled{0.7}{\textwidth}{
               &&\nVoid{out}\\
   \nLfor{lfor}&&\nRPA{po1}\\
   \nCut{c5}   &&\simplenet{ax1}{\Pi}\\
   \simplenet[11mm]{rev}{\RevW[w]}
               &&\nLlin{appl1}\\
   \nCut{c6}   &&             &&\nRMtens{couple}\\
   \simplenet[11mm]{dtow}{\DtoW_l[s]}
               &&             &\simplenet[7mm]{zero}{\ZD}&
\nAx{aaa}&\simplenet[31mm]{sigma}{\Lift{k_{\varepsilon}}{(k-k_{\varepsilon})}[\srntolall{u_\varepsilon}{l}]}\\
               &&             &&\nLtens{lt1} &\nLtens{lt2}\\
               &&             &&\nCut{c1}    &\nCut{c2}\\
               &&             &&\simplenet{contr1}{\DDg{0}_{l}[s]}&\simplenet{dupl1}{\DDg{k}_{l}[s]}\\
               &&\nLPA{p1}    &&\nLPA{p2}&\nLPA{p3}\\
               &&\nLlin{appl2}&\simplenet{rocA}{\texttt{Step}_1}&\nCut{c4}\\
               &&\nLlin{appl3}&\simplenet{rocB}{\texttt{Step}_0}&\simplenet[15mm]{coe1}{\CoerceD^{1}_l[s]}\\
    \nVoid{inw}&&             &&\nVoid{i1}&\nVoid{i2}
    \ncline{po1}{out} \nbput{\S^{k+1}\ND}
    \ncline{ax1}{po1} \nbput{\S^{k}\ND}
    \ncline{appl1}{ax1} \nbput{A}
    \ncline{p1}{appl1} \nbput{A\lin A}
    \ncangle[angleA=90,angleB=0]{couple}{appl1}
      \nbput{A}
    \ncangle[angleA=90,angleB=180]{zero}{couple}
      \naput{\ND}
    \ncangle[angleA=90,angleB=0]{sigma}{couple}
      \nbput{\pas^{k}\ND}
    \ncline{p2}{contr1}
    \ncline{p3}{dupl1}
    \ncline{contr1}{c1}
    \ncline{c1}{lt1}
      \nbput{(\ND)^2}
    \ncline{lt1}{aaa}
    \ncline{aaa}{couple} \naput{\ND}
    \nccurve[angleA=40,angleB=220]{lt1}{sigma}
    \ncline{dupl1}{c2}
    \ncline{c2}{lt2}
      \nbput{(\pas^{k}\ND)^2}
   \nccurve[angleA=160,angleB=-60,linestyle=none]{-}{lt2}{couple}  \ncput[npos=0.6]{\nId{idfloat1}}
   \nccurve[angleA=160,angleB=-60]{lt2}{idfloat1}
   \ncline{idfloat1}{couple} \nbput{\pas^k\ND}
    \ncline{lt2}{sigma}
    \ncline{i1}{coe1} \nbput[npos=0.3]\ND
    \ncline{coe1}{c4} \nbput{\pas\ND}
    \ncline{c4}{p2}
    \ncline{i2}{p3}   \nbput[npos=0.2]{\pas^{k+1}\ND}
    \ncangle[linewidth=1.5pt,linestyle=dotted,angleA=180,angleB=180]{-}{p1}{po1}
    \ncline[linewidth=1.5pt,linestyle=dotted]{-}{p1}{p2}
    \ncline[linewidth=1.5pt,linestyle=dotted]{-}{p2}{p3}
    \ncangle[linewidth=1.5pt,linestyle=dotted,angleA=0,angleB=0,armB=15mm]{-}{po1}{p3}
    \ncline{appl2}{p1}
    \ncline{appl3}{appl2}
    \ncangles[angleA=0,angleB=-90,armA=5mm]{lfor}{appl3}
    \ncline{c5}{lfor}
      \nbput{\NC}
    \ncline{rev}{c5}
    \ncline{c6}{rev}
      \nbput{\NC}
    \ncline{dtow}{c6}
    \ncline{inw}{dtow}
      \nbput{\ND}
    \ncangles[angleA=90,angleB=45]{rocA}{appl2}  \naput[npos=.5]{\ocs(A\lin A)}
    \ncangles[angleA=90,angleB=45]{rocB}{appl3}  \naput[npos=.5]{\ocs(A\lin A)}
  }
 } 
 \end{minipage}
 \begin{minipage}{.4\textwidth}
 {
  \begin{center}
  where
   {\tiny
     $$A = \underbrace{\ND\otimes\ldots\otimes\ND}_{1+\numnor{n}}
      \otimes
      \underbrace{\pas^{k}\ND\otimes\ldots\otimes\pas^{k}\ND}_{\numsafe{s}+1}$$
    }
   and \texttt{Step}${}_i$, with $i\in{0,1}$, is:
  \end{center}
 }
 {
  \proofnetpagescaled{0.7}{\textwidth}{
                         &\nVoid{out}\\
                         &\nROC{rocA} \\
                         &\nRlin{lambdaA}\\
                         &\nRMtens{rtensA}\\
    \simplenet[15mm]{s0}{\SD_i[s]}&   \nId{idc}     &\simplenet[36mm]{thetaA}{\Lift{k_i}{(k-k_i)}[\srntolall{u_i}{\ob{t}{l}}]}\\
    \nLtens{lt1}         &\nLtens{lt2}    &\nLtens{lt3}\\
    \nCut{cut1}          &\nCut{cut2}     &\nCut{cut3}\\
    \simplenet[15mm]{contrS}{\DDg{0}_{\ob{t}{l}}[s]}
                         &\simplenet[15mm]{contr2}{\DDg{0}_{\ob{t}{l}}[s]}
                                          &\simplenet[15mm]{dupl2}{\DDg{k}_{\ob{t}{l}}[s]}\\
                         &\nLMtens{ltensA}&\\
    \ncline{rocA}{out}  \nbput{\ocs(A\lin A)}
    \ncloop[linewidth=1.5pt,loopsize=72mm,armA=30mm,armB=50mm,angleA=180,angleB=0]{-}{rocA}{rocA}
    \ncldownup[armA=26.5mm,armB=3mm]{lambdaA}{ltensA}
    \ncline{lambdaA}{rocA}
      \nbput{A\lin A}
    \ncline{rtensA}{lambdaA}
       \nbput{A}
    \ncangle[angleA=180,angleB=-90]{ltensA}{contrS}
    \ncline{ltensA}{contr2}
    \ncline{ltensA}{dupl2}
    \ncangle[angleA=0,angleB=-90, offsetB=-35pt]{ltensA}{thetaA}
      \nbput[npos=1.5]{\pas^k\ND}
    \ncangle[angleA=90,angleB=180]{s0}{rtensA}
       \naput{\ND}
    \ncline{contrS}{cut1}
    \ncline{cut1}{lt1}
       \naput{\ND^2}
    \ncline{lt1}{s0} \naput{\ND}
    \ncline{lt1}{thetaA}
    \ncline{contr2}{cut2}
    \ncline{cut2}{lt2}
       \naput{\ND^2}
    \ncline{idc}{rtensA} \naput{\ND}
    \ncline{lt2}{idc}
    \nccurve[angleA=45,angleB=220]{lt2}{thetaA}
    \ncline{dupl2}{cut3}
    \ncline{cut3}{lt3}
       \naput{(\pas^{k}\ND)^2}
    \nccurve[angleA=170,angleB=-80,linestyle=none]{-}{lt3}{rtensA}  \ncput[npos=0.7]{\nId{idfloat2}}
    \nccurve[angleA=170,angleB=-70]{lt3}{idfloat2}
    \ncline{idfloat2}{rtensA}
       \nbput{\pas^{k}\ND}
    \ncline{lt3}{thetaA}
    \ncangle[angleA=90,angleB=0,offsetB=1mm]{thetaA}{rtensA}
      \nbput[npos=0.5]{\pas^{k}\ND}
  }
 } 
 \end{minipage}
\end{tabular}
}
\begin{definition}[\titoloblocchi{Representing a term by a \pn}]
  Let $t$ be in $\SRN^{\numnor{n};\numsafe{s}}$,  $l\in\N$, and
  $\Pi \,\triangleright\, \vec{\ND}^{\numnor{n}}$ $\vec{(\S^k\ND)}^{\numsafe{s}}\vdash \S^k\ND$,
  for some $k\in\N$.
  Then, $\Pi$ $k$-\emph{simulates $t$ with $l$-bounded inputs} if,
  for every pair of vectors of natural numbers
  $\vec{x}^{\numnor{n}}, \vec{y}^{\numsafe{s}}$,
  such that $|\vec{x}^{\numnor{n}}|, |\vec{y}^{\numsafe{s}}|\leq l$,
  the \pn\ we get by plugging $\srntolall{x_1}{l},\ldots,\srntolall{x_{\numnor{n}}}{l},
                              \S^k\srntolall{y_1}{l},\ldots,\S^k\srntolall{y_{\numsafe{s}}}{l}$ into the inputs of $\Pi$,
			      in the natural way,
  normalizes to $\S^k\srntolall{z}{l}$, whenever $z$ is the result of $t(\vec{x}^{\numnor{n}};\vec{y}^{\numsafe{s}})$.
\end{definition}
\begin{proposition}[\titoloblocchi{\SRN\ embeds into \LALL}]
\label{proposition:tempname simulates SRN}
 Let $l\geq 0$, and $t\in\SRN^{\numnor{n};\numsafe{s}}$. Then,
 $\srntolall{t}{l}$ $k$-simulates $t$ with $l$-bounded inputs.
 Moreover, (i) $k\leq \degreeR{t}\cdot 2^{\degreeC{t}}$, and (ii)
$\size{\srntolall{t}{l}}$ is $O\left(l^{\degree{p_t}^{(\degreeC{t}+\degreeR{t})}}\right)$, namely a polynomial in $l$.
\end{proposition}
The statement holds by induction on $t$, using the definition of $\srntolallg$,
the size of every \pn\ that $\srntolallg$ generates, the definitions of $\nb{t}{\cdot}$, and $\ob{t}{\cdot}$, together with
Facts~\ref{fact:nb is a polynomial}, \ref{fact:ob bounds the output length}, and \ref{fact:SRN-numerals}.
\condinc{
 \input{finfun-into-mulal}
}{
}
\adaptedsection{Conclusions and further works}
\label{section:Conlcusions and further work}
We have shown that the compatibility between the \emph{predicative analysis} over recursive functions that \SRN\ encodes, and the proof theoretical \emph{stratification}, that regulates the complexity of some Light Logic that characterize \PTIME, can be improved, provided that
(i) the stratification we find in Light Linear Logic and \LAL\ relaxes to boxes with ``fuzzy'' border, as in \MLfour\ or \LALL, and
(ii) we move to a representation of words \emph{alternative to the standard one}, able both to represent the basic functions of \SRN\ in constant time, and to exploit the independence of the cut elimination complexity from the logical complexity of the \formulae\ in a \pn.
\par
As a consequence, every term $t$ of \SRN\ maps to a family $\lang \srntolall{t}{l}\rang_{l\in\N}$ of \pn s in \LALL, where $\srntolall{t}{l}$ simulates $t$ with inputs \emph{at most as long as} $l$. The number of paragraph modalities in the type of the conclusion of $\srntolall{t}{l}$ depends on the structural complexity of $t$.
The size of $\srntolall{t}{l}$ is a polynomial in $l$ whose degree depends on the degree of the characteristic polynomial of $t$ and on the structural complexity of $t$.
\par
As an example, the following program, which returns $y$ if $w\neq 0$ contains a digit '0' that is not the lowermost digit, and $z$ otherwise, is in \SRN\ but not in \BCminus:
$$\begin{cases}
    g(0;y,z) &= z\hspace{5cm}(*)\\
    g(s_iw;y,z)&= h(w;y,z,g(w;y,z))\enspace ,
  \end{cases} $$
where $h(w;y,z,t) = \texttt{cond}(\;;w,y,t)$.
The embedding we propose gives the family of \pn s that implement it in \LALL, while, it is worth remarking, it is unknown how to represent $g(w;y,z)$ inside \LAL.
\par
Admittedly, the representation of a term of \SRN\ by a family of \pn s, rather than as a unique \pn, is not standard.
For example, one might be tempted to observe that every function with finite domain is the initial fragment of some polynomial time function, so \LALL\ represents \emph{every function with finite domain}.
Beware, however, that \emph{not every algorithm} is in \LALL.
For example, in analogy with \cite{Leivant93}, we show an algorithm $\texttt{exp}$ that cannot exist as a \pn\ of \LALL\ because it calculates a non-polytime function.
$\texttt{exp}$ will be defined using the traditional \emph{non-predicative} recursive schemes, so that it is not a program of \SRN.
$\texttt{exp}$ is defined as follows.
We know that the two programs $\texttt{concat}(x;y)$, which concatenates two strings of bits, and $\texttt{double}(x;\,)=\texttt{concat}(x;\pi^{1,0}_1(x;\,))$ belong to \SRN. Then, we can define the recursive function $\texttt{exp}$:
{
$$
      \texttt{exp}(\Zero;\,) = \Suco(\,;\pi^{1,0}_1(\Zero;\,))
      \hspace{15mm}
      \texttt{exp}(\Suci(\,;x);\,) = \texttt{double}(\texttt{exp}(x;\,);\,) \qquad (i\in\{0,1\})
$$}
The program $\texttt{exp}$ \emph{is not} in \SRN\ because of the position of the argument that drives the unfolding.
\emph{This reflects into} \LALL, where
$\srntolall{\texttt{concat}}{l}:\ND,\pas\ND\vdash\pas\ND$ and $\srntolall{\texttt{double}}{l}:\ND\vdash\pas\ND$ exist,
but $\srntolall{\texttt{double}}{l}$ cannot be iterated because of the form of its type. So, $\srntolall{\texttt{exp}}{l}$ cannot be defined as a \pn\ of \LALL\ using the constructions of this paper.
\par
We conclude by an example about how the approach ``\SRN\ as family-of-proofs'' we present here can be rewarding.
We consider the following program:
$$\begin{cases}
   \texttt{gt}(0,y)&=\texttt{ False}\\
   \texttt{gt}(\texttt{s}_ix,y)&=
       \texttt{ if } (y = 0)
       \texttt{ then True else } \texttt{gt}(x,\texttt{pred}(y)).
  \end{cases}
$$
The program \texttt{gt} is such that $\texttt{gt}(x,y)=\texttt{True}$ iff $|x|>|y|$.
It has a recursive definition more liberal than the primitive recursion scheme, as the recursive call of \texttt{gt} applies a function on the parameter $y$ that does not drive the unfolding.
Namely, \texttt{gt} incorporates a \emph{double iteration}.
Certainly, \texttt{gt} cannot exist in \SRN\ in the form here above.
Instead, \LALL\ admits to represent \texttt{gt} as follows:
\begin{align*}
 \texttt{gt}^{\NC\lin\ND\lin\B}&=\lambda x^\NC.\lambda y^\ND.\pi_2\left(x\{\ND\otimes\B\}(\texttt{step})(y\otimes\FF)\right)\\
 \texttt{step}^{{\ND\otimes\B}\lin{\ND\otimes\B}}&=\lambda s^\ND\otimes b^\B.b{\{\ND\otimes\B\}}(0\otimes\TT)\left(s(0\otimes\TT)(\lambda x^\B.\lambda y^\ND.y\otimes\FF)\right)
\end{align*}
where $\B,\TT,\FF$ are at page~\pageref{def:booleans}.
The existence of \texttt{gt} in \LALL\ implies the existence of a family
$\lang\texttt{ord}_l\rang_{l\in\N}$ of \pn s. Every  $\texttt{ord}_l$ takes two Scott words with \emph{at most $l$ bits}, and gives them back sorted according to their length.
The definition of every $\texttt{ord}_l$ is a \pn\ of \LALL\ that we compactly write as a $\lambda$-term:
\begin{align*}
\texttt{ord}_l^{{\ND\otimes\ND}\lin{\ND\otimes\ND}}&=
\lambda x^\ND\otimes y^\ND.
(\lambda x_1^\ND\otimes x_2^\ND.
 \lambda y_1^\ND\otimes y_2^\ND.
      \texttt{BtoB}
      \left(\texttt{gt}\left(\DtoW_l[x_1]\right)y_1\right)
      (y_2\otimes x_2)
)(\DD_l\, x)(\DD_l\, y)
\\
 \texttt{BtoB}^{\B\lin{\mathbf B}}&=\lambda x^\B.x\{\mathbf B\}(\Lambda\gamma.\lambda w^\gamma\otimes z^\gamma.w\otimes z)(\Lambda\gamma.\lambda w^\gamma\otimes z^\gamma.z\otimes w)
\end{align*}
where
$\mathbf B=\forall\gamma.(\gamma\otimes\gamma)\lin(\gamma\otimes\gamma)$
is a linear version of the booleans, $\DtoW_l^{\ND\lin\NC}$ is at page~\pageref{def:DtoW}, and the \emph{safe duplication} $\DD_l^{\ND\lin\ND\otimes\ND}$ is at page \pageref{def:DD}.
\complexfigure
{Insertion sort for Scott Words no longer than $l$.}
{fig:insertion-sort}
{\small
\begin{align*}
 L(X) &= \forall\alpha.!(X\lin\alpha\lin\alpha)\lin\S(\alpha\lin\alpha)\qquad\mbox{is the type of lists},\\
 \emptyset^{L(X)} &=
 \lambda c^{!(X\lin\alpha\lin\alpha)}.\lambda z^\alpha.z\\
 \texttt{sort}_l^{L(\ND)\lin\S L(\ND)} &=
 \lambda t^{L(\ND)}.t\{L(\ND)\}(\texttt{insert}_l)\ \emptyset^{L(\ND)}\\
 \texttt{insert}_l^{\ND\lin L(\ND)\lin L(\ND)} &=
        \lambda n^\ND.\lambda t^{L(\ND)}.
        \lambda c^{!(\ND\lin\alpha\lin\alpha)}.
        \lambda z^\alpha.\\
        &\qquad
        (\lambda x^\ND\otimes y^\alpha.c\,x\,y) ((t\,\texttt{putTop[c]}_l)(\CoerceD^{1}_l[n]\otimes z))\\
 \texttt{putTop}[c]_l^{\ND\lin\ND\otimes\alpha\lin\ND\otimes\alpha}&=
        \lambda a^\ND.
        \lambda b^\ND\otimes t^\alpha.
        (\lambda i^\ND\otimes\lambda j^\ND.i\otimes c\,j\,t)
        (\texttt{ord}_l\, a\otimes b)\\
\end{align*}
}
\par
Given $\texttt{ord}_l$, we can write a family of insertion sorts that sort lists of Scott Words as much long as $l$. Figure~\ref{fig:insertion-sort} describes one element of the family.
We warn the reader about the syntax we use. It does not perfectly adhere to the one in Figure~\ref{fig:linear-lambda-terms}, but \pn s would consume too much space.
The effort to move from the terms in Figure~\ref{fig:insertion-sort} to the \pn s of \LALL\ they represent should be a reasonably simple exercise.
We observe that $\texttt{putTop}_l$ is a linear algorithm that manipulates only the \emph{head}
of a given list. Instead, $\texttt{insert}_l$ takes a number and a sorted list, and puts the number at the correct position of the list, so to preserve the sorting.
While performing an iteration, $\texttt{insert}_l$ does \emph{not} add any paragraph $\S$ in front of the type of the output. The reason is that it exploits the general scheme that allows to write a perfectly linear \emph{predecessor} on Church numerals in the $\lambda$-calculus \cite{Roversi:1999-CSL}.
Finally, $\texttt{sort}_l$, iterates $\texttt{insert}_l$ in the usual way, thus adding a $\S$ in front of its output type.

\paragraph{Future lines of research.}
We must say that the representation of \SRN\ as a family of \pn s of \LALL\ that we present in this work has been an alternative approach to the standard one, which would explore the relations between \SRN\ and stratification by mapping a single term of \SRN\ into a single \pn.
That standard approach has been developed in \cite{RoversiV08,RoversiVercelli-FOPARA09,Vercelli10}. Those works make some progress as compared to \cite{MurawskiOng00}. This means that they identify a Light Logic that strictly contains \LAL, and which allows to represent a strict extension of \BCminus. However, the whole \SRN\  still escapes any full representation inside a Light Logic. So, it has been natural to look for an alternative approach; and this brought us to this work.
\par
Naturally enough, future work is about ``integrating'' both \emph{by level technology} and \emph{multimodality}. Multimodality is in the framework \MS\ developed in the previously cited works \cite{RoversiV08,RoversiVercelli-FOPARA09,Vercelli10}.
The conjecture is that the two technologies together may lead to a more refined proof theoretical representation of the principles underpinning the definition of \SRN, and of the predicative analysis it encodes, possibly increasing the set of algorithms that we can represent inside Light Logics.

\bibliographystyle{eptcs} 
\bibliography{lal-and-mu}

\end{document}